\documentclass{aa}  

\usepackage{graphicx}
\usepackage{txfonts}
\usepackage{multirow}
\begin{document} 

   \title {Fine structure and long duration of a flare coronal X-ray source with RHESSI and SDO/AIA data}
   
   \subtitle{}
   
   \author{S. Ko{\l}oma\'nski
        \inst{1}
        \and
        T. Mrozek
        \inst{1,2} 
        \and 
        E. Chmielewska
        \inst{1}}

   \institute{Astronomical Institute of University of Wroc{\l}aw, Poland\\
           \email{kolomans@astro.uni.wroc.pl}\\
           \email{chmielewska@astro.uni.wroc.pl}
           \and
           Solar Physics Division, Space Research Centre, Polish Academy of Sciences, Poland\\
           \email{mrozek@astro.uni.wroc.pl}}

   \date{Received ... / accepted ...}

   \abstract
{Coronal X-ray sources (CXSs) are phenomenon very often occurring in solar flares regardless of a flare size, duration or power. The nature of the sources was difficult to uncover for many years. It seems that at last, combining data from \textit{RHESSI} and \textit{SDO}/AIA, there is a unprecedented possibility to 'look inside' CXSs and to answer the questions about their formation, evolution and structure.}
{We present a study of a CXS of the SOL2011-10-22T11:10 long-duration flare observed simultaneously with \textit{RHESSI} and \textit{SDO}/AIA. We focus our attention on the following questions: What was responsible for the CXS presence and long duration? Was there any fine structure in the CXS?}
{The AIA instrument delivers high quality images in various EUV filters. \textit{RHESSI} data can be used to reconstruct images in X-rays and to perform imaging spectroscopy. Such a complementary data enables to study a relation between the CXS and structures observed in EUV during the decay phase of the flare.} 
{X-ray emission recorded by \textit{RHESSI} during the decay phase of the flare came from about 10~MK hot CXS. The source was observable for 5 hours. This long presence of the source could be supported by magnetic reconnection ongoing during the decay phase. Supra-arcade downflows, which are considered to be a manifestation of magnetic reconnection, were observed at the same time as the CXS. The source was co-spatial with the part of the hot supra-arcade region that had the highest emission measure and simultaneously the temperature within the range of \textit{RHESSI} thermal-response. However, while the supra-arcade region was a dynamic region consisting of small-scale structures, the CXS seemed to be smooth, structureless. We run simulations using real and synthetic \textit{RHESSI} data, but we did not find any strong evidence that the CXS had any small-scale structure.}
{}

\keywords{Sun: corona -- Sun: flares -- Sun: X-rays, gamma rays -- Sun: UV radiation}


\maketitle

\section{Introduction}

High energy radiation from the Sun is highly non-uniform, confined to areas where hot plasma or non-thermal particles are present. During solar flares the patches of emission become far more localized and fewer. Energy contained in magnetic fields is in the course of a flare converted (via magnetic reconnection) in e.g. thermal and non-thermal energy of plasma which can be radiated, inter alia, in extreme ultraviolet and X-rays. These processes are confined to relatively small volumes but they are very effective and powerful. 

As the result, X-ray radiation of the Sun is dominated by small bright centers of two types: foot-point sources and coronal sources. The first type is placed in the lowest parts of magnetic arches i.e. at the chromosphere and the transition level. The second type is located high in the solar corona. Based on observations coronal X-ray sources (CXSs) can be divided into several classes due to their emission properties (thermal, non-thermal, mixed), location in a flare structure or acceleration mechanisms. For detailed discussion of different classes of CXSs observed in hard X-rays see \citet{krucker2008}. However, as it is pointed out by the authors, such a classification is not straightforward due to present day observational limitation. In our work we will focus on CXSs which are observed just above above warm ($\approx 1$~MK) post-flare loops and the acronym CXRs will be used throughout the paper in this particular sense. It is worth to notice that a term ``post-flare loops'' is considered by some authors as misnomer \citep{priest2000, svestka2007}. In fact those loops are visible when a flare is sill ongoing. Thus, instead of ``post-flare loops'' we will use a term ``post-reconnection flare loops'', PFLs.

CXSs are phenomenon very often occurring in solar flares regardless of a flare size, duration or power. They were discovered in 1970s in observations taken from the Skylab space station \citep[]{kahler1977}. Since that time the knowledge of CXSs has increased significantly due to the next generation of solar space instruments. Investigation of CXRs carried out by many authors led to many important conclusions \citep[e.g.][]{vorpahl1977, acton1992, doschek1995, feldman1995, doschek1996, jakimiec1998, white2002, jiang2006, kolomanski2011}:

\begin{itemize}
	\item CXSs are filled with hot ($\geq10$~MK) and relatively dense plasma ($10^{10}-10^{11}$~cm$^{-3}$)
	\item their emission is dominated by thermal emission, however sometimes a non-thermal component is also present
  \item physical parameters of plasma in CXSs (e.g. temperature, density) change smoothly with time
  \item continuous energy input must be present as the sources last longer than characteristic time of cooling of hot plasma
\end{itemize}

There is long-standing discussion concerning the fine structure of CXSs. \textit{Yohkoh} \citep[]{ogawara1991} observations showed CXSs as a coherent region of size of the order of $100-1000$~arcsec$^2$ (see e.g. \citet{pres2009}). Despite this relatively large size, CXSs were seen as diffuse without any small-scale structure. However, data from the SXT telescope \citep[]{tsuneta1991} have an angular resolution only 3.7~arcsec and a low thermal resolution (broad band filters). These characteristics may not be sufficient to answer the question about fine structure of CXSs. As a consequence it is hindered to answer other questions concerning formation of CXSs and their slow and gradual evolution.

{\em TRACE} \citep[]{handy1999}, launched in 1998, cast doubts on the diffuse nature of CXSs. This EUV telescope had higher spatial resolution than \textit{Yohkoh}/SXT. In {\em TRACE} images of solar flares sets of sharp loops are usually seen instead of diffuse coronal sources (see Fig.~2 in \citet{warren1999}). At first glance it would seem that the diffuse appearance of CXSs comes from the insufficient spatial resolution of SXT and that the sources are in fact composed of multitude of small elements, e.g. bright tops of filamentary loops. However, {\em TRACE} is the most sensitive to plasma at quite different temperatures ($1-2$~MK) than SXT ($\approx$ 10 MK). Thus, the hot plasma of CXSs is barely visible in {\em TRACE} images. 

Nevertheless, some {\em TRACE} observations of solar flares, especially in the 195~{\AA} band, reveal diffuse structures located above narrow PFLs. Location of these diffuse structures corresponds well with the {\em Yohkoh} CXSs (see e.g. Fig.~2 in \citet{warren1999}). Thus, the diffuse nature of CXSs cannot be explained only by the low spatial resolution. There is another instrumental factor that may blur an image of CXSs and veil the true nature of the sources. Plasma in solar flares is multithermal. A thermal response of the {\em TRACE} 195~{\AA} channel has two distinct maxima (see Fig.~3 in \citet{phillips2005}). The higher (in response) maximum comes from FeXII line, that forms at temperatures from 1 to 2~MK (``warm plasma''), whereas the second one is produced by FeXXIV line in temperatures from 10 to 30~MK (``hot plasma''). Due to different thermal widths of these two maxima, the cooler maximum selects much less emission elements (e.g. loops) from the multitude of elements with different temperatures than the hotter maximum. Thus, diffuse appearance of CXSs might be a result of blending of fine elements of very different temperature. It seems that, despite the very good spatial resolution, the {\em TRACE} thermal resolution for hot, flare plasma is insufficient to study CXSs. Moreover, TRACE response to hot plasma is almost two orders of magnitude lower than to warm plasma. Due to this fact, hot coronal sources are hard to observe in a presence of bright post-reconnection flare loops.

The EUV Imaging Spectrometer (EIS) \citep[]{culhane2007} is one of the three scientific instruments aboard {\em Hinode} \citep[]{kosugi2007}. EIS provides simultaneous observations in several spectral lines. The instrument can record emission of hot plasma in e.g. a band that includes the line of CaXVII at 192.82~{\AA}. The line forms at temperature $\approx4.5-7.5$~MK, i.e. it can partially reveal hot plasma that is present in a flare. Moreover, the thermal resolution in this line is higher than the {\em TRACE} thermal resolution for hot plasma. \citet{pres2009} analyzed EIS observations of the long-duration flare SOL2006-12-17T17:12. The authors reported that the flare emission was seen in the form of sharp filamentary loop structures in lines with formation temperatures below 3~MK (e.g. FeXII 195.12~{\AA}), as in typical {\em TRACE} images. In the Ca XVII image, there was an additional set of sharp loops with a distinct, diffuse CXS \citep[Fig.~5]{pres2009} - an image similar to that what was seen in the SXT data. Nevertheless, this observations and similar ones of EIS, cannot give us key information about the nature of CXSs. EIS images are obtained by rastering with a slit. This process takes significant amount of time. In the case of the SOL2006-12-17T17:12 flare the scanning through the CXS took about 15 minutes. It may be long enough to blur any details of the source. Thus, the EIS temporal resolution may be not appropriate in the case of CXSs. 

\textit{RHESSI} \citep[]{lin2002} observations offer another possibility to study nature of CXSs \citep[e.g.][]{jiang2006, vaananen2007, caspi2010}. The satellite enables us to obtain images and spectra of solar X-ray sources with a good angular (2.3~arcsec) and spectral (1~keV for imaging spectroscopy) resolution. \citet{kolomanski2011} analyzed three long-duration flares. The authors used \textit{RHESSI} imaging spectroscopy to obtain physical parameters of CXSs and to calculate the energy balance of the observed sources. One of the conclusions is that each CXS observed was smooth. No CXS was visible in the images reconstructed for the four narrowest grids (angular resolution $2.3-12$~arcsec). This suggests that either CXSs are diffusive with no internal structure or they are a superposition of small unresolved subsources with separation of the subsources smaller than the angular resolution of the finest \textit{RHESSI} grid. Thus, the question about the structure of CXSs was left without answer.

\textit{RHESSI} is an excellent instrument to observe coronal X-ray sources. It covers very wide range of radiation energy from 3~keV to 17~MeV with good resolution and high sensitivity. However it has some limitations. First, it is a Fourier imager thus, images are reconstructed based on recorded flux modulation additionally disturbed by noise. Second, limited dynamical range (usually 10:1) and lack of sensitivity to plasma colder than $\approx7$~MK make it difficult to relate directly CXSs to colder and/or fainter structures in flares. Therefore, a good idea is to supplement \textit{RHESSI} with additional data from an EUV directly imaging telescope.

Gallagher et al. (2002) used combined observations of \textit{RHESSI} and \textit{TRACE} to analyze the SOL2002-04-21T01:50 flare. Relation between EUV structures (\textit{TRACE}) and X-ray sources (\textit{RHESSI}) was investigated. The authors found that during the rise phase of the flare CXS was at the same or greater altitude in the corona as diffuse and hot emission recorded by \textit{TRACE} above warm PFLs. The CXS was observable also during the decay phase of the flare for 11 hours and it was always at higher altitude than the tops of PFLs. The authors note that there was no evidence of any fine structure of the CXS.

Some instrumental characteristics of \textit{TRACE}, especially, as mentioned above, relatively low sensitivity to hot plasma, limit analysis that can be performed on combined \textit{RHESSI} -- \textit{TRACE} data set. Fortunately, at present there is an instrument which can significantly improve the situation. The Atmospheric Imaging Assembly (AIA) onboard the \textit{SDO} satellite delivers high quality images in various EUV filters. The instrument has good angular, thermal and time resolution and is highly sensitive to hot plasma. Thermal responses of AIA and \textit{RHESSI} overlap in the range $\approx7-16$~MK allowing to study a relation between coronal X-ray sources and structures observed in EUV.

We present the study of a coronal source of the SOL2011-10-22T11:10 long-duration flare observed simultaneously with \textit{RHESSI} and \textit{SDO}/AIA. Using such complementary data we are able to study a relation between the CXS and EUV emission of the flare in greater detail than it was possible earlier. We focus on the following questions: What was responsible for the CXS presence and long duration? Was there any fine structure in the CXS? 

We selected a long-duration flare (LDE -- long-duration event) for our analysis and gave our attention mainly to its decay phase. The reasons are as follows. LDE flares are characterized by a very slow evolution, especially during the decay phase. Moreover, LDEs occur in large magnetic structures -- loops in such flares may reach the height of $10^5$~km. Any instrumental drawbacks like insufficient angular or temporal resolutions may be less limiting in the case of slowly evolving large-scale structures of LDEs. Basic characteristics of coronal X-ray sources, like slow evolution, smooth appearance and resistance, are exceptionally conspicuous and surprising during the decay phase of LDEs, where the sources may last for hours on end.

\section{Observations and data analysis}

\begin{figure}
\resizebox{\hsize}{!}{\includegraphics{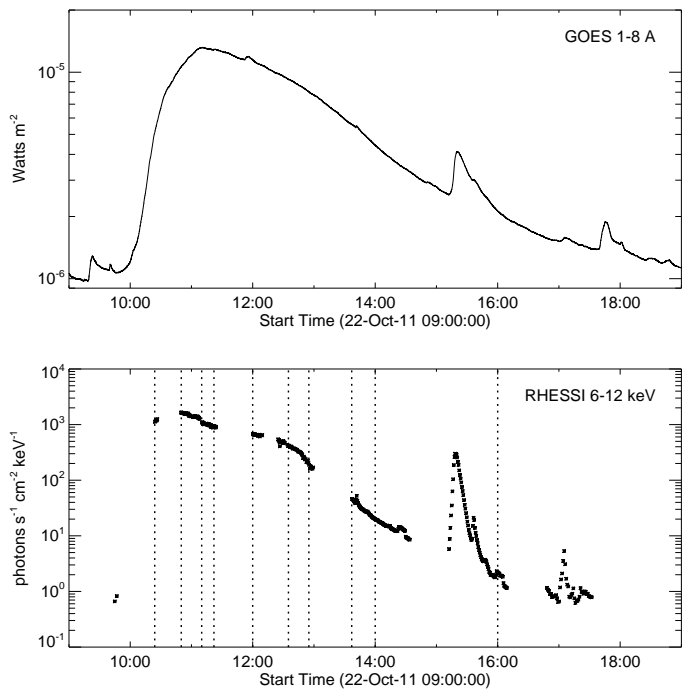}}
\caption{\textit{GOES} and \textit{RHESSI} light curves of the SOL2011-10-22T11:10 flare. Gaps in the \textit{RHESSI} curve are caused by the satellite nights and the South Atlantic Anomaly. Combined \textit{RHESSI} - AIA analysis presented in this paper was performed for 10 time indicated marked by the dotted lines. The length of the intervals is 16 to 180 seconds.}
\label{fig1}
\end{figure}

The SOL2011-10-22T11:10 LDE flare occurred in the active region NOAA~11314, close to the west solar limb. The \textit{GOES}/SEM (Space Environment Monitor, \citet{donnelly1977}) and \textit{RHESSI} light curves of the flare are presented in Fig.~\ref{fig1}. According to \textit{GOES} data the flare started at 10:00~UT. The rise phase lasted very long and the maximum of brightness in $1-8$~{\AA} band (M1.3) was reached at 11:10~UT. The solar X-ray flux returned to the pre-flare level at about 19:30~UT. The flare was a typical example of so-called long-duration event with the slow rise phase (sLDE) \citep[]{hudson2000, baksteslicka2011}. The data analyzed here were obtained by the Atmospheric Imaging Assembly (AIA) \citep[]{lemen2012} installed onboard the {\it Solar Dynamics Observatory (SDO)} \citep[]{pesnell2012} and by {\it Reuven Ramaty High Energy Solar Spectroscopic Imager (RHESSI)}. The SolarSoftWare (SSW, \citet[]{freeland1998}) system was used to analyze the data.

The {\it SDO}/AIA consists of a set of four 20~cm, normal-incidence telescopes. The field of view covers entire solar disk with $4096 \times 4096$ CCDs. AIA provides observations with high angular ($\approx$ 1.5~arcsec) and temporal ($\approx$ 10~s) resolutions in 10 bands including 7 EUV bands. The AIA observations cover the whole duration of the analyzed flare. For our study we selected six AIA bands, i.e. 94~{\AA}, 131~{\AA}, 171~{\AA}, 193~{\AA}, 211~{\AA} and 335~{\AA}. All six bands were used to calculate differential emission measure distributions. Two of these bands, 94~{\AA} and 131~{\AA}, are the best choice to study morphology and dynamics of hot plasma in regions of coronal sources. Each of these two bands has two narrow maxima in thermal response: warm ($\approx$ 1~MK) and hot ($\approx$ 10~MK), and both maxima are almost equal in sensitivity \citep[]{odwyer2010, boerner2012}. Thus, a hot component of flare emission is not overwhelmed by a warm component as it was in the case of the {\em TRACE} 195~{\AA} band. Moreover, the 94~{\AA} and 131~{\AA} bands have high thermal resolutions for hot maxima. The resolution is as good as for the warm maxima, and much higher than the resolution of {\em TRACE} 195~{\AA} band for hot plasma. 

{\it RHESSI} is a rotating Fourier imager with nine germanium detectors. Detectors are large ($7.1$~cm in diameter x $8.5$~cm in height) and cooled to about 75~K, thus the sensitivity of the instrument is great, and allows for investigation of very low solar flare fluxes that are observed during the decay phase of long-duration flares. The instrument was launched in February 2002 which results in lower sensitivity after several years of observations. However, there is a possibility for a partial restoration of the instrument's capabilities thanks to annealing\footnote{http://sprg.ssl.berkeley.edu/\~{}tohban/nuggets/?page= article\&article\_id=69}. The last annealing, before the October 2011, was performed in March 2010, therefore we may expect that, for the analyzed event, detectors are in a good condition and still provide us with valuable scientific data. 

The \textit{RHESSI} light curve of the analyzed flare is shown in Fig.~\ref{fig1}. The light curve is not continuous due to the satellite nights and the South Atlantic Anomaly. Thus, the flare is not fully covered by \textit{RHESSI} observations. Dotted lines in the figure indicate 10 short time intervals, for which we reconstructed {\it RHESSI} images. There are currently several algorithms available for \textit{RHESSI} image reconstruction. Images for our analysis were reconstructed using PIXON algorithm \citep{pina1993}. The energy resolution chosen for reconstruction was usually $1-2$~keV. In a few cases wider energy ranges ($4-6$~keV) were necessary because of very low signal (late decay phase, energy above $10$~keV). We chose PIXON algorithm, which is generally considered as very reliable technique in image reconstruction\footnote{http://hesperia.gsfc.nasa.gov/rhessi3/software/imaging-software/image-algorithm-summary}. Moreover, based on our experience, we think that PIXON algorithm combined with the grid selection method (described below) is the best option to study coronal sources. However, it should be kept in mind that reconstruction of images based on flux modulation recorded by the \textit{RHESSI} detectors is complicated. The recorded flux is disturbed by noise of solar and non-solar origin. In result we deal with an ill-posed inverse problem and uncertainty of synthesized images. In order to verify reliability of the obtained PIXON images we reconstructed for some of the 10 selected time intervals also control images using other algorithms and the grid selection method. These control images are similar to PIXON images except higher energies where recorded flux is low. In such a conditions other used reconstruction algorithms usually failed while PIXON algorithm gave good quality images. We used images to perform also imaging spectroscopy. For this purpose we need as wide energy coverage as possible. PIXON images meet this requirement most satisfactorily.

X-ray emission observed several hours after a flare maximum is extremely weak, and usually X-ray images cannot be reconstructed with standard parameters. We used a method of grid selection described in \citet{kolomanski2011}. Coronal X-ray sources observed by the authors were large and diffuse. In such a case, the reconstruction method does not converge with ``standard'' grid selection, i.e. a set consisting of grids No. 3-9, producing more or less noisy distribution of small sources. The situation may be improved. We should keep in mind that a signal from a large source is modulated only by grids which have resolution worse than the actual source size. The decision, which grids should be used, is made on a basis of Back Projection (BP) images reconstructed for individual grids. Only grids showing clear modulation of source signal are selected for the final image reconstruction. In the case of the analyzed flare the modulation was not present in grids No. 1-5 in any selected time interval. Thus, we used only grids with lower resolution, starting from grid No. 6, to reconstruct images for our study. Later in this paper we test how a grid selection method behaves in the case where an CXS consists of several small-scale sources.

Due to low X-ray emission during the decay phase of flares longer time intervals are required to reconstruct reliable \textit{RHESSI} images, i.e. images with high enough photon statistics. We used time intervals from 16 to 180 seconds long. Slow temporal evolution of the analyzed flare, including slow changes of the CXS position, enables to use such a long time intervals. Length of the intervals were selected accordingly to keep number of counts in the range $10^3 - 10^4$ per one image. Furthermore, we used data phase stacking, i.e. combining data on the basis of \textit{RHESSI} roll angle. The stacking improves statistics and enables imaging over long time intervals.

\begin{figure*}
\centering
\includegraphics[width=17cm]{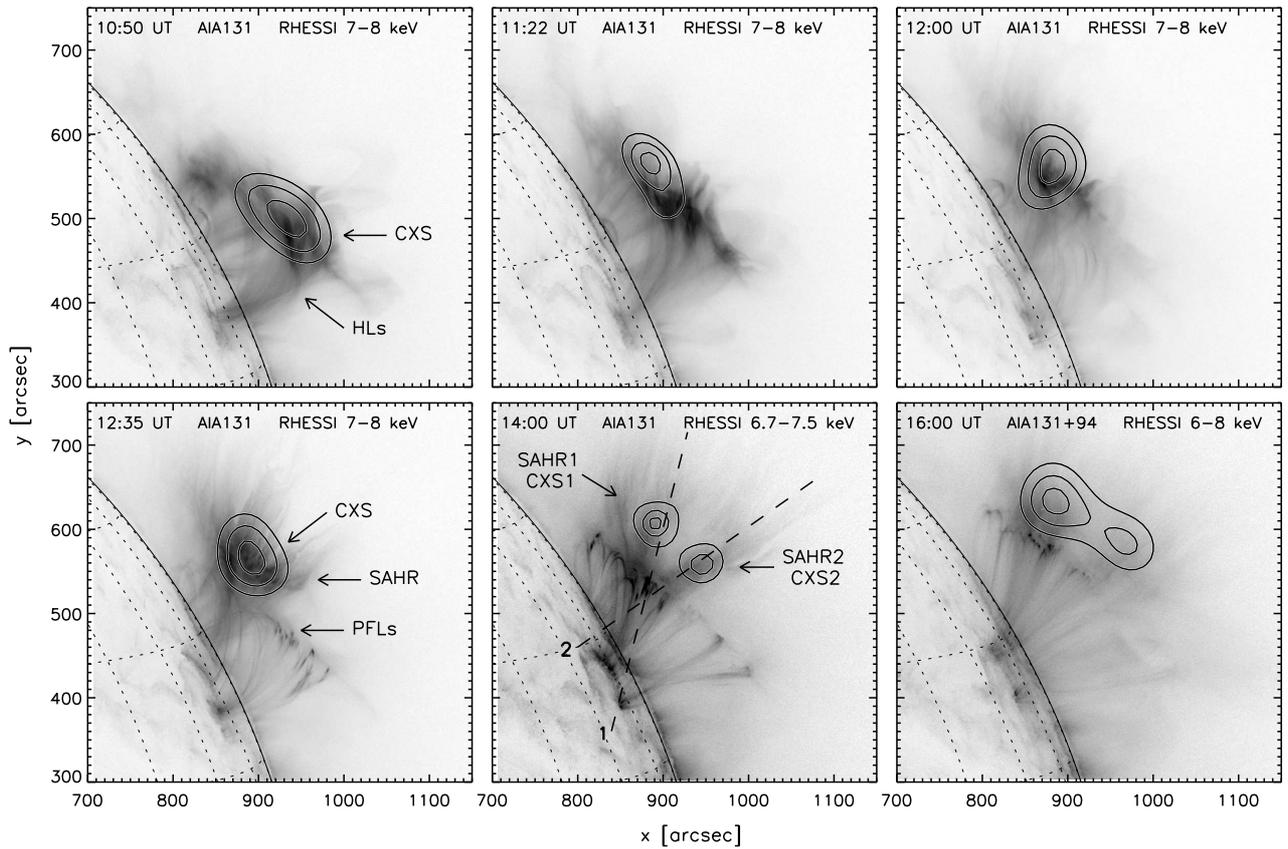}
\caption{\textit{SDO}/AIA images illustrating the evolution of the SOL2011-10-22T11:10 flare. Contours show the coronal X-ray source (CXS) observed with \textit{RHESSI} in the energies given in each image. In the images the flare hot loops (HLs), the supra-arcade hot region (SAHR) with supra-arcade downflows (SADs), and the post-reconnection flare loops (PFLs) are visible. The dashed lines mark the cuts (1 and 2) for which difference dynamic maps were constructed (see Fig.~\ref{fig7}).}
\label{fig2}
\end{figure*}

We selected 10 time intervals for a combined \textit{RHESSI} - AIA analysis. The intervals are indicated in Fig.~\ref{fig1} by dotted lines. A set of AIA images with \textit{RHESSI} contours is shown in Fig.~\ref{fig2}. With help of the figure we can take a look at the flare evolution. The SOL2011-10-22T11:10 event began with an eruption at about 10:00~UT. A set of bright and hot loops (HLs, $T\approx10$~MK) appeared just after the start of the eruption. The loops were visible only in the AIA bands with significant sensitivity to plasma at temperatures around 10~MK or more (i.e. 94~{\AA} and 131~{\AA} bands). The apexes of the HLs were their brightest parts. Just after the maximum of the flare (11:10~UT according to \textit{GOES} data) the HLs started to fade due to plasma cooling. They reappeared later, roughly at 12:00~UT, as an arcade of warm post-reconnection flare loops (PFLs, $T \approx 1-2$~MK). Simultaneously with the PFLs, a bright supra-arcade hot region (SAHR) began to be visible. As in the case of the HLs, the SAHR was also distinctly visible only in the AIA 94~{\AA} and 131~{\AA} bands, i.e. it had temperature $\approx10$~MK.

The SAHR showed clearly a well known phenomenon -- supra-arcade downflows (SADs) \citep[]{mckenzie1999, mckenzie2000}, i.e. dark features moving downwards in the SAHR towards the PFLs. The SADs were also visible in the 94~{\AA} band. During the flare decay phase PFLs got higher and the SAHR moved slowly upwards and became fainter. Since about 13:00~UT the SAHR is seen divided into two parts (SAHR1, SAHR2) separated by the fainter region between. The SAHR1 and SAHR2 were visible as areas of convergence of SADs, giving them the appearance of two hands. The SAHR was visible in the 131~{\AA} band until almost 17~UT. The 94~{\AA} band is more sensitive to hot than to warm plasma. Thus, images in this band were dominated by hotter plasma of the SAHR and the PFLs were fainter than in the 131~{\AA} band.

The images reconstructed using \textit{RHESSI} data (contours in Fig.~\ref{fig2}) show only coronal X-ray source (CXS). Despite the fact that the whole structure of the flare was visible (no occulting by the solar disk), there were no detectable X-ray sources at the loops' footpoints. The position of the CXS followed the brightest part of hot component in the AIA 131~{\AA} images for most of the duration of the flare (see Fig.~\ref{fig2}). Before the flare maximum the CXS was located at or very near the bright apexes of the HLs. Such a situation lasted for about an hour until the HLs cooled down. However the CXS continued its existence because a new hot region emerged, i.e. the SAHR. After the maximum the CXS was located just above the PFLs, in the SAHR. Before about 13:00~UT there was only one X-ray source and after 13:00~UT -- there were two (CXS1 and CXS2). The division took place at the same time as the division of the SAHR in the AIA 131~{\AA} band. The last image of the CXS, which we were able to reconstruct, is for 16:00~UT. No more than one hour later the SAHR became practically invisible.

Good energy resolution of the reconstructed images enabled us to perform an imaging spectroscopy of the observed CXS. The imaging spectroscopy has a few advantages in comparison with a spectroscopic analysis of the entire solar signal, without a spatial resolution. First, during an image reconstruction the background is naturally subtracted. The problem of appropriately subtracted background is severe in the case of long-duration flares, which are characterized by very low signals during the decay phase. Second, during the long decay phase, other flares may occur and mask the original emission from an LDE. The flare observed around 15:20~UT (see Fig.~\ref{fig1}), which was in another active region, far from the analyzed flare, is a good example. In such a case the imaging spectroscopy allows to distinguish the actual emission from the flare despite the fact that the additional emission from another flare was present.

The physical parameters of the observed coronal X-ray source obtained from imaging spectroscopy are shown in Table~\ref{tbl-1}. This analysis could be done for the eight first time intervals. For the last two intervals there was not enough images in different energy ranges (too low signal above 10~keV) to fit the spectrum. The CXS spectrum could be fitted with two thermal components. No non-thermal component was detected. The temperatures of the components were $8-9$ and $13-23$~MK respectively, which corresponds well to the hot structures observed by AIA, i.e. the HLs and SAHR.

\begin{table*}
\caption{The geometrical and physical parameters of the observed coronal X-ray source obtained from \textit{RHESSI} data (images and imaging spectroscopy).}
\label{tbl-1}
\centering
\begin{tabular}{c c c c c c}
\hline\hline
time  &  fitted   & temperature & emission         & cross-section  & altitude  \\
      & component &             &  measure         & area           & of centroid  \\
$\left[ \rm{UT} \right]$  &     & [MK] & [$10^{49}$~cm$^{-3}$] & [arcsec$^{2}$] & [$10^{3}$~km]\\
\hline
\multirow{2}{*}{10:24} & low  & 8.6  & 1.66   & \multirow{2}{*}{7980} & \multirow{2}{*}{61.0} \\
      & high & 23.1 & 0.0020 &        \\
\hline
\multirow{2}{*}{10:50} & low  & 7.5  & 6.42   & \multirow{2}{*}{6740} & \multirow{2}{*}{67.3} \\
      & high & 16.4 & 0.010  &         &   \\
\hline
\multirow{2}{*}{11:10} & low  & 6.7  & 7.67   & \multirow{2}{*}{8770} & \multirow{2}{*}{59.7} \\
      & high & 13.7 & 0.031  &         &   \\
\hline
\multirow{2}{*}{11:22} & low  & 8.7  & 1.64   & \multirow{2}{*}{4840} & \multirow{2}{*}{62.8} \\
      & high & 17.8 & 0.0022 &         &   \\
\hline
\multirow{2}{*}{12:00} & low  & 8.6  & 1.22   & \multirow{2}{*}{4820} & \multirow{2}{*}{60.4} \\
      & high & 15.3 & 0.0062 &         &   \\
\hline
\multirow{2}{*}{12:35} & low  & 8.6  & 0.88   & \multirow{2}{*}{6510} & \multirow{2}{*}{68.9} \\
      & high & 13.9 & 0.0022 &         &   \\
\hline
\multirow{2}{*}{12:55} & low  & 8.5  & 0.14   & \multirow{2}{*}{9050} & \multirow{2}{*}{76.4} \\
      & high & n/a   & n/a   &         &   \\
\hline
\multirow{2}{*}{13:37} & low  & 7.8  & 0.071  & \multirow{2}{*}{2410} & \multirow{2}{*}{85.6} \\
      & high & 12.6 & 0.0042 &         &   \\
\hline
\end{tabular}

\tablefoot{Parameters for the low and high temperature component, that were fitted to the observed spectra, are given. Combined parameters $T$ and $EM$ are given for both parts of the CXS visible at 13:37~UT (CXS1 and CXS2). The projected area and the centroid altitude of the CXS were determined from the \textit{RHESSI} images in the energy range $5.5-7.5$~keV. The CXS was defined by intensity isoline 0.5 with respect to the brightest pixel of the CXS.}

\end{table*}

\section{Results}

\subsection{CXS structure}

The coronal X-ray source observed by \textit{RHESSI} during the SOL2011-10-22T11:10 flare showed typical characteristics as described in the previous papers on coronal sources of long-duration flares \citep[e.g.][]{kolomanski2011}. \textit{RHESSI} images reconstructed with the use of the mentioned grid selection method show large and smooth CXS, without any internal structure. The position of the CXS during the decay phase was coincident with the supra-arcade hot region seen in AIA images. However, unlike the CXS, the SAHR was not structureless. The SAHR consisted of many brighter and fainter small-scale areas (see e.g. Fig.~\ref{fig2}, 12:35~UT image). Thus, we should answer the question whether the smooth appearance of the CXS is real or is caused by characteristics of reconstruction methods used and the instrument itself (Fourier imager).

We decided to check sensitivity of the grid selection method to a scenario in which an CXS consists of several small-scale sources. Such an approach is motivated by the AIA observations of the analyzed flare. As previously mentioned, the observations show many EUV structures spatially correlated with X-ray emission.

\begin{figure*}
\centering
\includegraphics[width=17cm]{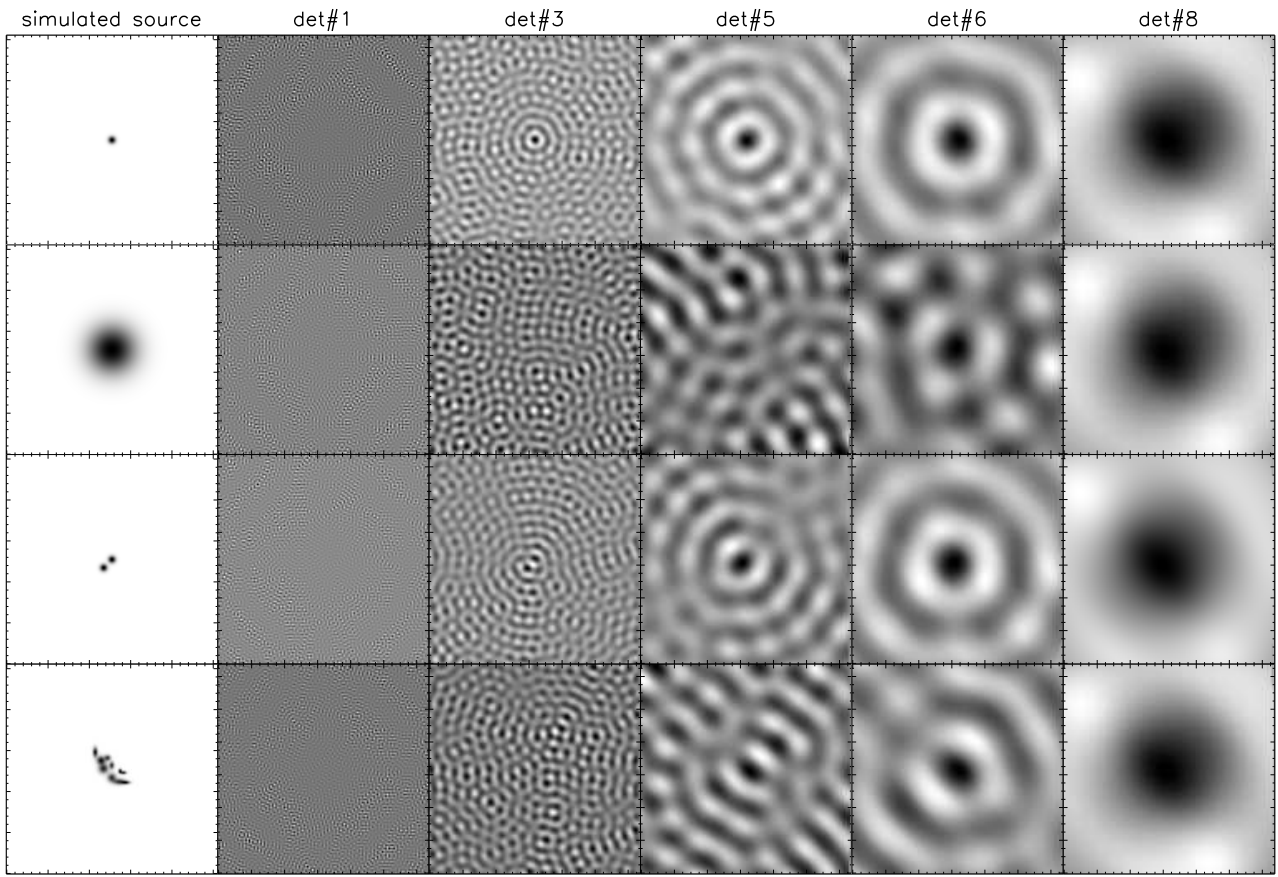}
\caption{Left column: simulated distributions of X-ray sources. Remaining columns: single-grid, back projected images of simulated sources. A single large source, and distribution of small sources in elliptical area are similar in terms of modulation seen in single-grid images. Namely, the modulation of signal is visible starting from the grid No.~6 which suggest that source has diameter between 20-30~arcsec. A single small, and two small sources are both visible starting from the grid No.~3 which is the finest grid with resolution worse than the actual size of these sources.}
\label{fig3}
\end{figure*}

If X-ray emission comes from a number of small sources, occupying a relatively small area, then signal modulation is not present in finer grids, and, as a consequence, we misinterpret it as one large source. The example of such a case is presented in Fig.~\ref{fig3}. The first column shows several of simulated scenarios: a single small (3 arc sec) source, a single large (20 arc sec) source, two small sources spaced at 10 arc sec, and ten small sources (3-7 arc sec) spread on elliptical area. Back Projection (BP) single grid images reconstructed for the scenarios are presented in the remaining columns of Fig.~\ref{fig3}. It is clear that for a single source or double sources the grid selection method works properly. Namely, the small source (3 arc sec) gives modulation visible on single grid images starting from grid the No.~3 (resolution 6.79 arcsec). The size of the larger source is comparable with the resolution of the grid No.~5. Therefore, we observe clear modulation starting from the grid No.~6 which has resolution 35.27 arcsec. The double source consisting of small sources separated by 10 arc sec is barely visible in the grid No.~3 image, but both sources are noticeable. The multiple-source scenario in single grid images is similar to the single large source scenario. There is a weak periodicity visible in the grid No.~5 image which is connected to elliptical shape of the area covered by small sources, but, generally, a fine structure cannot be recognized in single-grid images. Thus, it is possible that the large and smooth CXS of the analyzed flare had small-scale structure, similar to that seen in the AIA images.

Simulations of \textit{RHESSI} sources were performed with the use of standard software available in the SSW package. It allows to include typical background count rates for a given segment of detector. We tried several levels of background and photon count rates to verify how much it may affect our results. Signal-to-noise ratios from 3 to 1000 (count rates $10^1-10^6$) were simulated. We concluded that noise do not affect reconstruction results significantly even for S/N ratios as low as 3. It is obvious if we remember that background is almost not modulated by \textit{RHESSI} rotation while solar flare flux is \citep{hurford2002}. In real observations we are able to register weak 4~s modulation of X-ray flux reflected by Earth but the level of such signal is far below solar flare signals. The range of count rates tested in our numerical simulations covers the range of count statistics in images reconstructed for real \textit{RHESSI} data ($10^3 - 10^4$ counts per one image).

\begin{figure*}
\centering
\includegraphics[width=17cm]{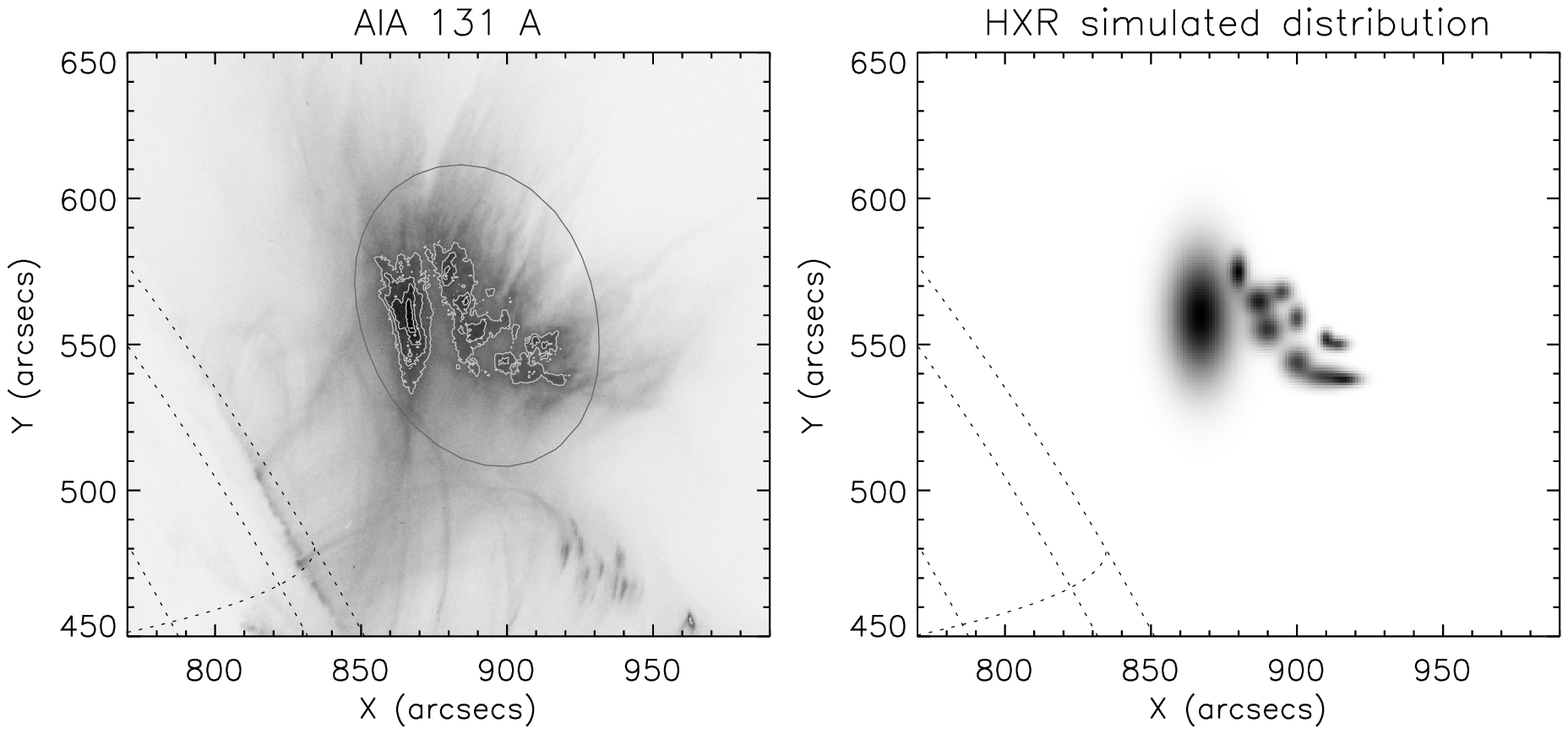}
\caption{Left: \textit{SDO}/AIA131 image with the intensity contours (0.6, 0.7, 0.8, and 0.9 of brightest pixel). Moreover the coronal X-ray source in the energy 6-7~keV is shown by 0.5 contour (in gray). The EUV emission within the CXS was not uniform, there were many small bright regions. Right: synthetic CXS used in our simulation. The synthetic CXS consists of a set of eleven small X-ray sources with sizes, locations, and relative intensities similar to EUV brightest regions visible in the AIA 131~{\AA} image within the CXS.}
\label{fig4}
\end{figure*}

We ran the following simulation to check the possibility that analyzed flare had small-scale structure. An AIA 131~{\AA} image was used as a proxy of a possible distribution of a fine structure of the CXS. In Fig.~\ref{fig4} (left panel) we present the AIA 131~{\AA} image taken at 12:35:33~UT. The white contours are drawn for EUV emission with levels of 0.6, 0.7, 0.8, and 0.9 of the brightest pixel. The gray contour represents 0.5 level of the maximum of the \textit{RHESSI} source (CXS) reconstructed for 2-minute long time interval (12:35-12:37~UT). We assumed that the CXS might have the fine structure similar to the structure observed in the AIA 131~{\AA} image. Such an assumption can be justified by the fact that hot plasma visible in the AIA image (T$\approx$10~MK) can be also observed by \textit{RHESSI}.

\begin{figure*}
\centering
\includegraphics[width=17cm]{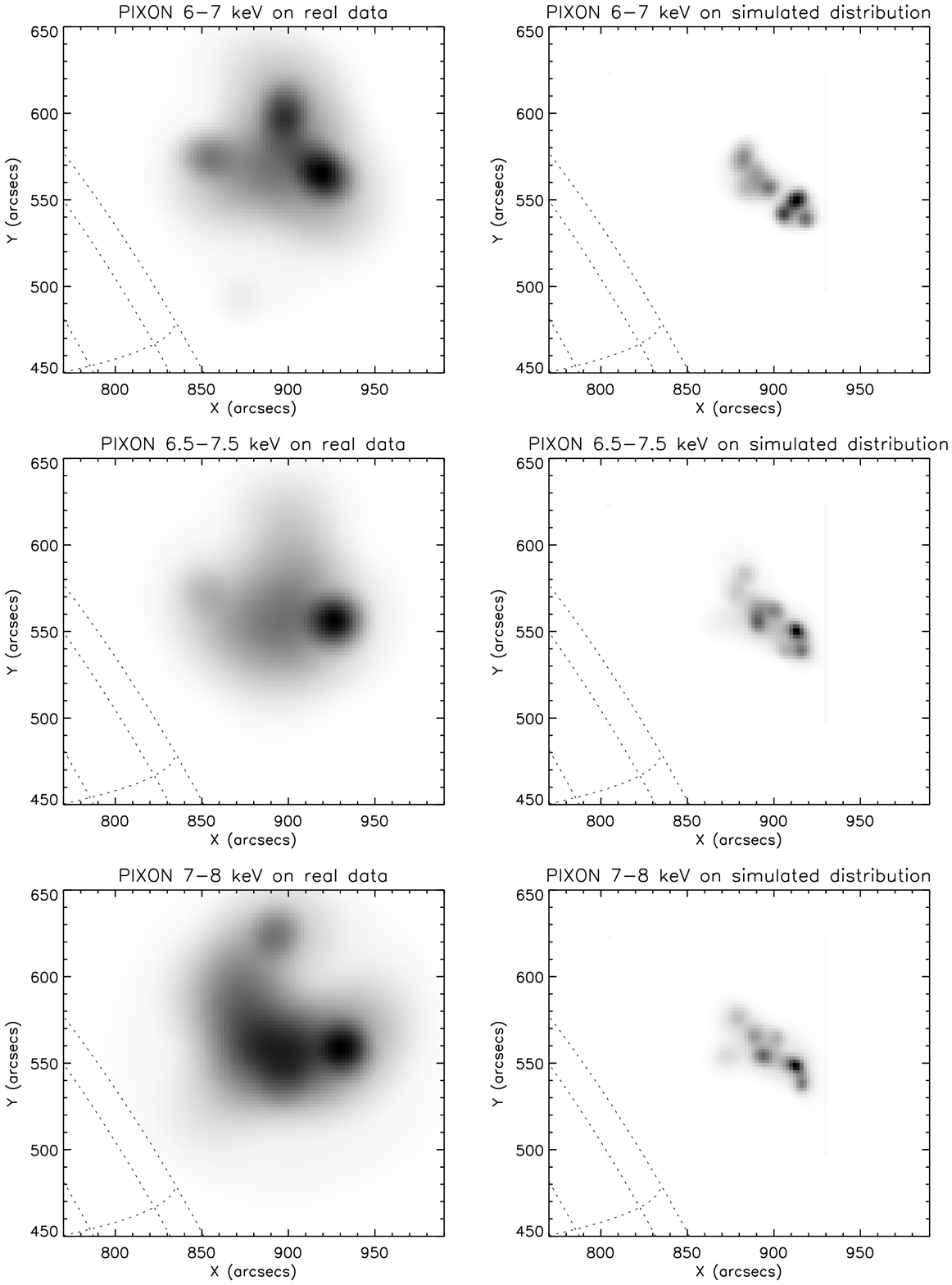}
\caption{Left column: PIXON images of the real X-ray emission of the SOL2011-10-22T11:10 flare reconstructed for the 6-7, 6.5-7.5, and 7-8~keV energy bands, and for the time interval 12:35-12:37~UT. Right column: PIXON images reconstructed for the same energy bands and grids but for simulated distribution of X-ray sources (synthetic CXS) shown in Fig.~\ref{fig4} (the right image).}
\label{fig5}
\end{figure*}

We defined a set of eleven small X-ray sources with sizes, locations, and relative intensities similar to EUV sources visible in the AIA 131~{\AA} image within the CXS. Ratio of brightness of the this small sources is $\leq$1.5:1. The ratio is smaller than \textit{RHESSI} dynamic range (typically 10:1, \citet{lin2002}). Thus, all the sources should be visible in \textit{RHESSI} images simultaneously. The simulated set of sources is presented in Fig.~\ref{fig4}, the right panel. Having such a synthetic CXS consisting of sub-sources we reconstructed its \textit{RHESSI} images using the PIXON algorithm with the grids No. 3-6, 8, and 9. Although the reconstructed images show a set of small sub-sources but their sizes, locations, and relative intensities are different from the ones in the assumed model. Moreover images reconstructed for slightly different energy ranges show a significantly different set of sub-sources (see images in the right column of Fig.~\ref{fig5}). Thus, we can conclude that the sub-sources and overall fine structure in reconstructed images are not real. The fine structure of the synthetic CXS was not possible to recover.

The fact that we do not see the largest of the simulated sub-sources constitutes an unexpected result of this test. We performed several other tests in which we tried other scenarios with one larger sub-source and various combinations of smaller ones. Even for a two-source scenario (one large, and one small) we obtained an image with the smaller sub-source only. The larger sub-source is completely invisible, it was lost during the reconstruction process. This unexpected result was not mentioned in previous papers testing performance of various image reconstruction algorithms \citep[]{aschwanden2004, schmahl2007, dennis2009}. From the interpretation point of view such a result has serious consequences. It is possible that in reconstructed images of X-ray emission based on real observations we may not see large, diffuse sources when smaller sources are present near by.

From this paper's point of view, such artificial disappearance of a larger source can help us to answer the question about fine structure of the CXS of the analyzed flare. The presence of a large, diffuse CXS in reconstructed images of the flare may suggest that there are no small sources, namely a fine structure is less likely. We checked this supposition by comparing images reconstructed for real data to images reconstructed for our simulated distribution of sub-sources. The real images were obtained for the same energy intervals (6-7, 6.5-7.5, and 7-8~keV) and the same set of grids (No. 3-6, 8, and 9) as simulated ones. Keeping in mind the results of our single-grid test, we reconstructed these images with the use of fine grids, down to the grid No. 3, despite lack of modulation in those grids. Using overlapping intervals we wanted to see if any of the reconstructed sources is visible in the same location in energy-neighboring images.

The resulting images are presented in Fig.~\ref{fig5} (the left column), and compared to the images reconstructed for the synthetic CXS (the right column of the same figure). The sources reconstructed for the real data show almost no repeatability, i.e. the images reconstructed for adjacent energy ranges are not similar, suggesting that a fine structure is absent. In order to increase the certainty of this result we reconstructed additional images with the same parameters and real data using five other algorithms available in \textit{RHESSI} software. Outcome was similar as in the PIXON images, fine structure was not visible. There is only one source that is present in all three PIXON images around position $x=925, y=560$. However, we doubt that the source is real due to the following reasons. The source is not in the same position in energy-neighboring images. Its centroid in 6-7~keV is shifted with respect to 7-8~keV by a value similar to the source size. Moreover, there is no bright structure in AIA images nor in differential emission measure maps (see subsection 3.3) around $x=925, y=560$. Furthermore, the source is not present in any image reconstructed with BP, Clean, EM, MEM NJIT, UV Smooth algorithms.

However, the same situation is observed in the case of the images reconstructed for the synthetic CXS which consists of sub-sources. Nevertheless, the images reconstructed for the real and simulated data differ. In the case of the synthetic CXS the PIXON algorithm managed, to some extent, to reconstruct its fine structure, i.e. the sizes of the obtained sub-sources are comparable to their simulated input values, despite different positions. For the real data we did not observe such a behavior. The images reveal big sources that are significantly larger than the resolution of the finest used grid (No.~3). We also reconstructed PIXON images for real data adding the grid No.~1 to the set of grids. The resulting images are almost the same as without the grid No.~1.  We interpret this result as a lack of fine structure in the observed source. However, it should be noticed that a similar result may be produced in the case of many small sources with fast changes of intensity (fast in comparison with time of integration used for imaging).

Such a model of an CXS was presented by e.g. \citet{longcope2010}. In the model the CXS consists of a set of transient, randomly occurring sub-sources. In the case of the flare analyzed by the authors, a typical lifetime of each sub-source was about 8 seconds (about two orders of magnitude shorter than the duration of the coronal source of the flare). Thus, if \textit{RHESSI} images are reconstructed for time intervals of e.g. tens of seconds or even several minutes, CXSs in such images will be a superposition of many transient sub-sources.

In the case of the SOL2011-10-22T11:10 flare we used times of integration as long as 2-4 minutes to reconstruct \textit{RHESSI} images. Much shorter intervals should be used to check if the CXS is a superposition of transient, randomly occurring sub-sources. However, this can not be done due to the low level of signal. Instead of \textit{RHESSI} images we can use AIA images. As an example we compare again the \textit{RHESSI} image reconstructed for time interval 12:35-12:37~UT with the AIA images. As mentioned, there were EUV sources visible in the AIA 131~{\AA} images within the area of the X-ray CXS (see Fig.~\ref{fig4}). If the large diffuse CXS was the superposition of transient sub-sources this should be visible in the AIA images. We compared all ten AIA 131~{\AA} images taken during the 12:35-12:37~UT interval and found no evidence of such transient sub-sources. All the visible EUV sources are quite stable in their location and brightness. No new EUV source appeared and no existing source extinguished. Thus, if the CXS was superposition of transient sub-sources it can not be detected by neither \textit{RHESSI} nor AIA. 

However, a small-scale structure of the CXS can not be decisively excluded. Firstly, spectra of CXS emission suggest the presence of two thermal components. The components may origin from two thermodynamically different types of areas within the CXS. Secondly, AIA images combined with the \textit{RHESSI} data show that the CXS was co-aligned with the supra-arcade hot region which was full of small-scale structures. One of these small-scale structures were supra-arcade downflows. In the next subsection we analyze SADs dynamics and possible physical connection between the downflows and the coronal X-ray source.

\subsection{CXS and SADs}

We carefully prepared and studied difference dynamic maps (space-time maps) to verify this possible connection. SADs are rather weak when compared to other parts of a flare. Thus, we took the AIA images in 193~{\AA}, 94~{\AA}, and 131~{\AA} bands, and constructed running difference images for each band separately. Then we extracted a narrow cut from each running difference image and stack them in time sequence \citep[]{sheeley2004}. After this operation we got an image with time running along x-axis and distance measured along the extracted path on y-axis. On such difference dynamic (DD) maps all the moving structures (e.g. SADs) are visible as bright and dark tracks.
 
\begin{figure*}
\centering
\includegraphics[width=17cm]{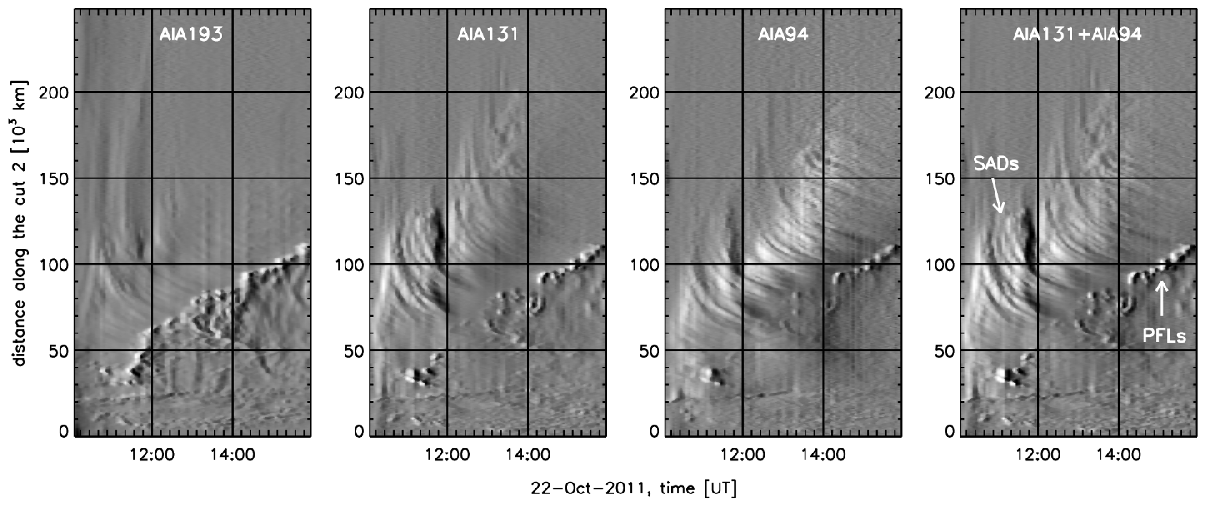}
\caption{Comparison of the difference dynamic (DD) maps obtained from the \textit{SDO}/AIA images made in the AIA 193~{\AA}, 131~{\AA} and 94~{\AA} bands. The last DD map is the combination of the 131~{\AA} and 94~{\AA} maps. Each DD map is made for the same time period and the cut 2 (see Fig.~\ref{fig2}). Supra-arcade downflows (SADs) are visible as bright and dark tracks moving downwards. Dark and bright smudges, visible below the SADs, are the tops of rising post-reconnection flare loops (PFLs).}
\label{fig6}
\end{figure*}

The SADs were very faint in the 193~{\AA} DD maps. Fortunately, they are distinctly visible in the 94~{\AA} and 131~{\AA} ones (see Fig.~\ref{fig6}). The maps show clearly the difference in the thermal response of the AIA bands and the multithermal character of the flare emission. In each of the three DD maps the post-reconnection flare loops are visible in similar position because all the three bands have the (local) maximum of the thermal response around $1$~MK. The PFLs got higher with the average velocity of $3-4$~km~s$^{-1}$. On the other hand, at a given time the SAD tracks were visible at higher position in the 131~{\AA} band than in the 94~{\AA} band. Both bands have the significant sensitivity to hot flare plasma. However, the 131~{\AA} band's maximal sensitivity for hot plasma is around $11-12$~MK while for the 94~{\AA} band it is around $7-8$~MK. The DD maps in these two bands give complementary information about hot plasma of the SAHR and therefore about the SADs which were visible as voids in the SDHR. Due to this we decided to combine the 94~{\AA} and the 131~{\AA} maps into one 94+131 DD map (see Fig.~\ref{fig6}). It is worth to notice that the third sensitive to hot plasma band, i.e. the 193~{\AA} band, does not provide additional information about the SADs. Although the 193~{\AA} band's maximal sensitivity for hot plasma is around $17-19$~MK, the SAD tracks are visible at roughly the same altitude as on the 131~{\AA} map at a given time.

\begin{figure*}
\centering
\includegraphics[width=17cm]{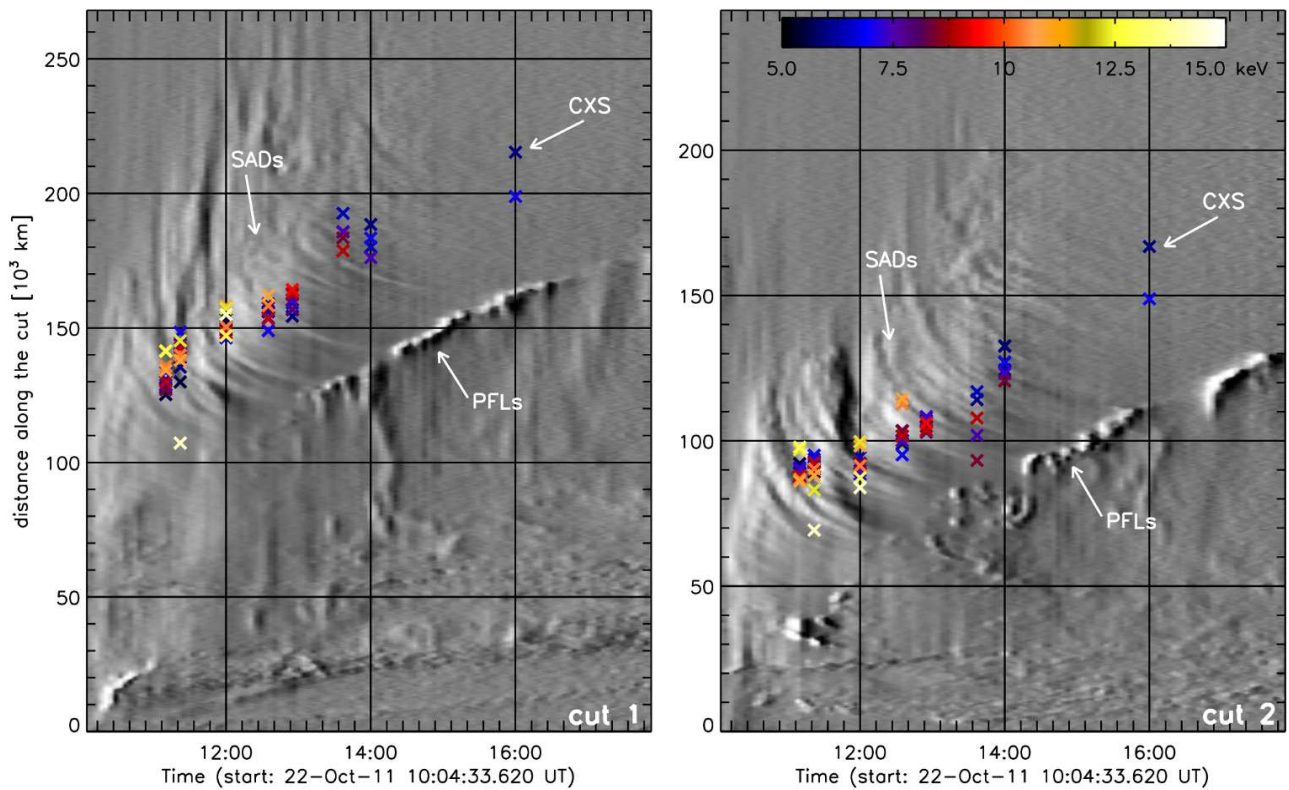}
\caption{Difference dynamic (DD) maps obtained from the combination of the \textit{SDO}/AIA images made in the AIA 131~{\AA} and 94~{\AA} bands. The DD maps are made for two cuts, which start at the footpoints of the PFLs and run up to follow the position of the CXS before 13:00~UT (one source) and positions of two sub-CXSs that came into existence due to division of the CXS (see Fig.~\ref{fig2}). The left DD map includes the northern sub-CXS (LTS1), the right one -- the southern sub-CXS (LTS2). The positions of centroids of the CXSs determined on the {\it RHESSI} images are marked with crosses. Color of the crosses show energy for which the \textit{RHESSI} images were reconstructed. See the electronic edition of the Journal for a color version of this figure.}
\label{fig7}
\end{figure*}

Fig.~\ref{fig7} shows difference dynamic maps for two cuts. The cuts start at the footpoints of the PFLs and run up to follow the changing position of the CXS including its division into LTS1 and LTS2 (see Fig.~\ref{fig2}). Each of the two cuts follows up one of these sub-CXS. The positions of the CXS centroids estimated on the {\it RHESSI} images are marked in the DD maps. Colors code energy for which an image of the source was reconstructed. The CXS moved horizontally before the flare maximum (see two first images in Fig.~\ref{fig2}). Due to this the source is outside the selected cuts for two first time intervals for which we reconstructed the \textit{RHESSI} images. The position of the CXS are not shown in the figure for these time intervals.

The bright and dark tracks of the SADs reveal that the downflows were decelerated from almost $100$~km s$^{-1}$ to about $2$~km s$^{-1}$. These are typical values observed for SADs in many flares (see e.g. \citet{savage2011, warren2011, liu2013}). For detailed analysis of SADs of the SOL2011-10-22T11:10 flare see \cite{savage2012}. Velocities of the downflows estimated by us are similar to those obtained by \cite{savage2012}.

The coronal source was located at the altitude where the SADs slow down and accumulate above the PFLs. Constant and distinct stream of the SADs was present since the maximum of the flare for about four hours. After 15:00~UT the SADs became virtually invisible. Since the SADs were visible as voids in the supra-arcade hot region, their visibility during the late decay phase of the flare may be hindered by a very low brightness of the SAHR. Both, the SAHR and the CXS, vanished soon after the SADs, about 16--17~UT. Thus, a presence of the SAHR and the CXS coincides in time with the SADs. If the downflows are manifestation of magnetic reconnection as they are considered \citep{asai2004, khan2007} then this coincidence in time can explain the long-lasting coronal X-ray source. This hot source needs constant energy supply and the reconnection is a source of the energy. See Sect.~\ref{sec-dis}. for detailed discussion of this topic.

\subsection{CXS and SAHR thermal structure}

A comparison of \textit{RHESSI} and AIA images indicates that what the first instrument sees as CXS during the decay phase of the flare is a part of the supra-arcade hot region observed by the second one. Hence, a question arises. Why the CXS was observed in the lower part of the SAHR? To answer the question we need to determine the thermal structure of the flare above the PFLs. Firstly, we can look at the position of the CXS centroids marked in Fig.~\ref{fig7}. The relation the higher the energy the higher the altitude is not clearly shown by the centroids. Such a behavior may be caused by an absence of distinct vertical stratification of the temperature in the SAHR (namely, the higher the altitude the higher the temperature). Temperature distribution in the analyzed flare should be known to verify this supposition.

Using a proper set of AIA images such distribution in a form of differential emission measure (DEM) distribution can be prepared. In order to calculate the DEM in temperature range $\log T=5.5-7.5$, the six EUV bands of AIA are used, i.e. 131~{\AA} (peak of temperature response log T=7.05, \citet{odwyer2010}), 94~{\AA} (log T=6.85), 335~{\AA} (log T=6.45), 211~{\AA} (log T=6.30), 193~{\AA} (log T=6.20), and 171~{\AA} (log T= 5.85). The EUV band 304~{\AA} is optically thick and has a small response to flare-like temperatures, thus is disregarded. The input data are processed to the level 1.6 by deconvolving the point spread function with the \texttt{aia\_deconvolved\_richardsonlucy.pro} procedure in SSW package before data calibration by \texttt{aia\_prep.pro}. Checking of co-alignment of images is made by using \texttt{aia\_coalign\_test.pro} routine.

Determination of DEM is not straightforward. Observed intensities in e.g. broad band filters, constitute the convolution of the response function of an instrument and DEM, additionally disturbed by errors. In result we deal with an ill-posed inverse problem \citep[]{tikhonov1963, bertero1985, craig1986, schmitt1996, prato2006}. Thus, we can not be sure that an obtained solution is actual. In order to increase reliability of the obtained DEM it is recommended to use more than one method. For this reason we calculated DEM with the use of two methods.

The first method (hereafter method-1) applies the \texttt{xrt\_dem\_iterative2.pro} routine in SSW package \citep[]{weber2004, golub2004} modified for use with the AIA filters (see details in the appendix of \citet{cheng2012}). This is an iterative forward fitting method in which the algorithm minimizes differences between the fitted and the observed intensities measured in the six EUV AIA bands. Errors in the method-1 are estimated using the Monte Carlo (MC) approach. Each of 100 MC simulations is disturbed by a randomly drawn noise within an uncertainty in the observed flux in each AIA band. The uncertainty is computed using \texttt{aia\_bp\_estimate\_error.pro} procedure in SSW package.

The second used method, regularized inversion technique (hereafter method-2), was introduced by \citet{hannah2012}. The method is computationally very fast and calculates also both the vertical (emission measure) and horizontal (temperature) error bars. In both methods errors are higher for (a) the very low signal in six used AIA bands and for (b) temperatures close to edges of range of the temperature response functions of these AIA bands (1-20~MK).

\begin{figure*}
\centering
\includegraphics[width=17cm]{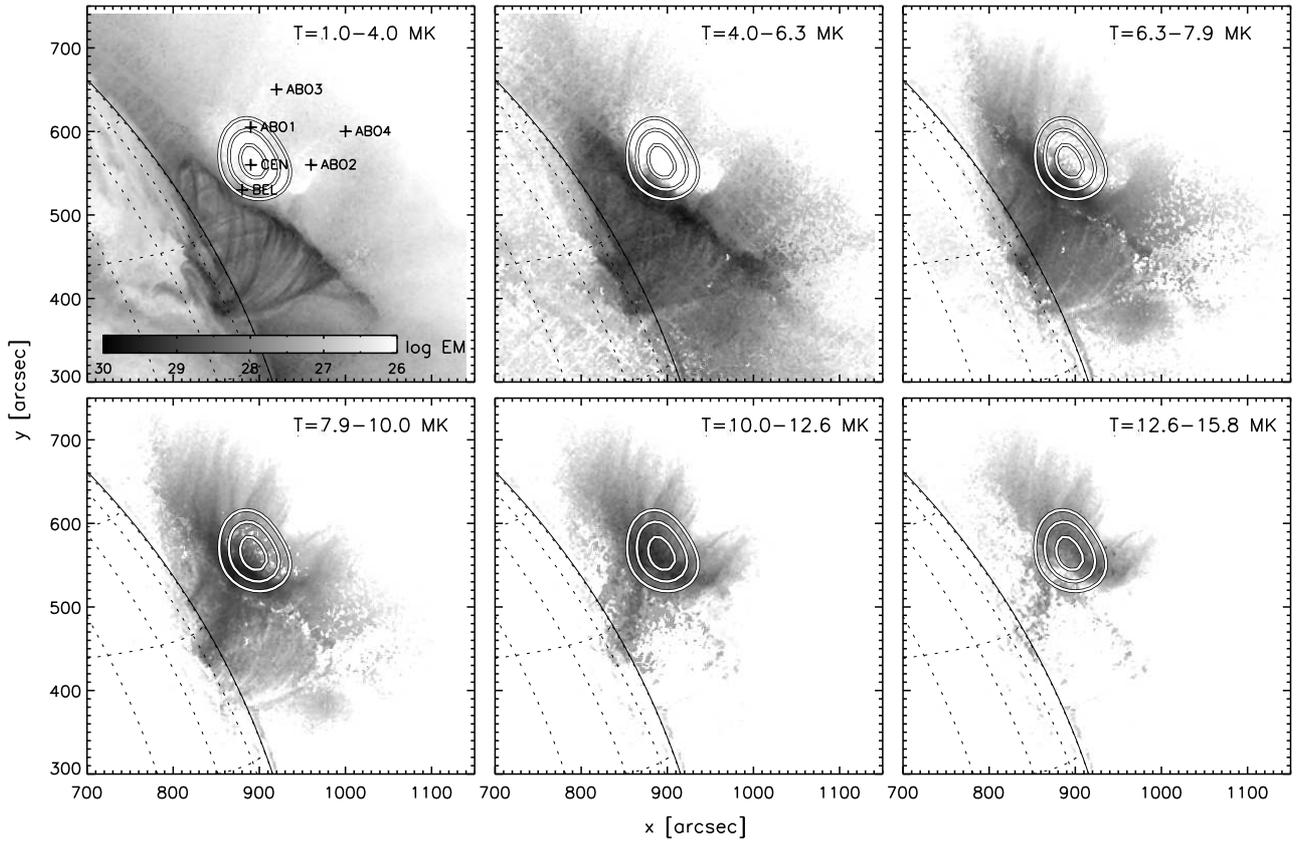}
\caption{DEM maps of the SOL2011-10-22T11:10 flare at 12:35~UT calculated with the use of \textit{SDO}/AIA images for six temperature ranges. Contours show the coronal X-ray source observed with \textit{RHESSI} in the energy range 7-8~keV. Points marked in the first map show regions for which DEM distributions are shown in Fig.~\ref{fig9}.}
\label{fig8}
\end{figure*}

Again, for a detailed analysis we took AIA observations taken at 12:35~UT, i.e. the same time for which we analyzed a structure of the CXS. Using the method-1 we prepared DEM maps for the selected time in the following way. Firstly, AIA images were resized down by factor of two, i.e. each pixel in resized image is sum of a box 2 by 2 original pixels. This reduces spatial resolution but simultaneously reduces also an influence of noise on DEM. Secondly, DEM was calculated for each new pixel based on intensities measured in six EUV bands. The resulting emission measure maps in different temperature ranges are shown in Fig.~\ref{fig8}.

In each map a DEM value is shown by a gray scale for subsequent temperature ranges. Contours show the CXS position as observed with \textit{RHESSI} in the energy range 7-8~keV. The lower temperature ranges (1-4~MK) are dominated by emission of the PFLs. There is no significant emission from the CXS at these temperatures. A distinct void is visible in this and the next temperature range (4.0-6.3~MK) in the location of the SAHR and CXS. Plasma of these two regions emits intensively in the temperature above 8~MK. DEM maps show a mild vertical gradient of a temperature within the SAHR (and CXS). The brightest emission at 6.3-7.9~MK and 7.9-10.0~MK comes from the lowest part of the SAHR (and CXS). At a bit higher altitude, closer to the centroid of the CXS, the majority of plasma at the temperature of 10-12.6~MK is concentrated. The brightest emission at the temperature 12.6-15.8~MK comes from the upper part of the CXS. Thus, the CXS is made of the brightest areas of the emission at temperature above 8~MK, arranged in the following order: the higher the altitude the higher the temperature. We can not tell, based on this maps, whether this trend continues above the CXS.

\begin{figure*}
\centering
\includegraphics[width=17cm]{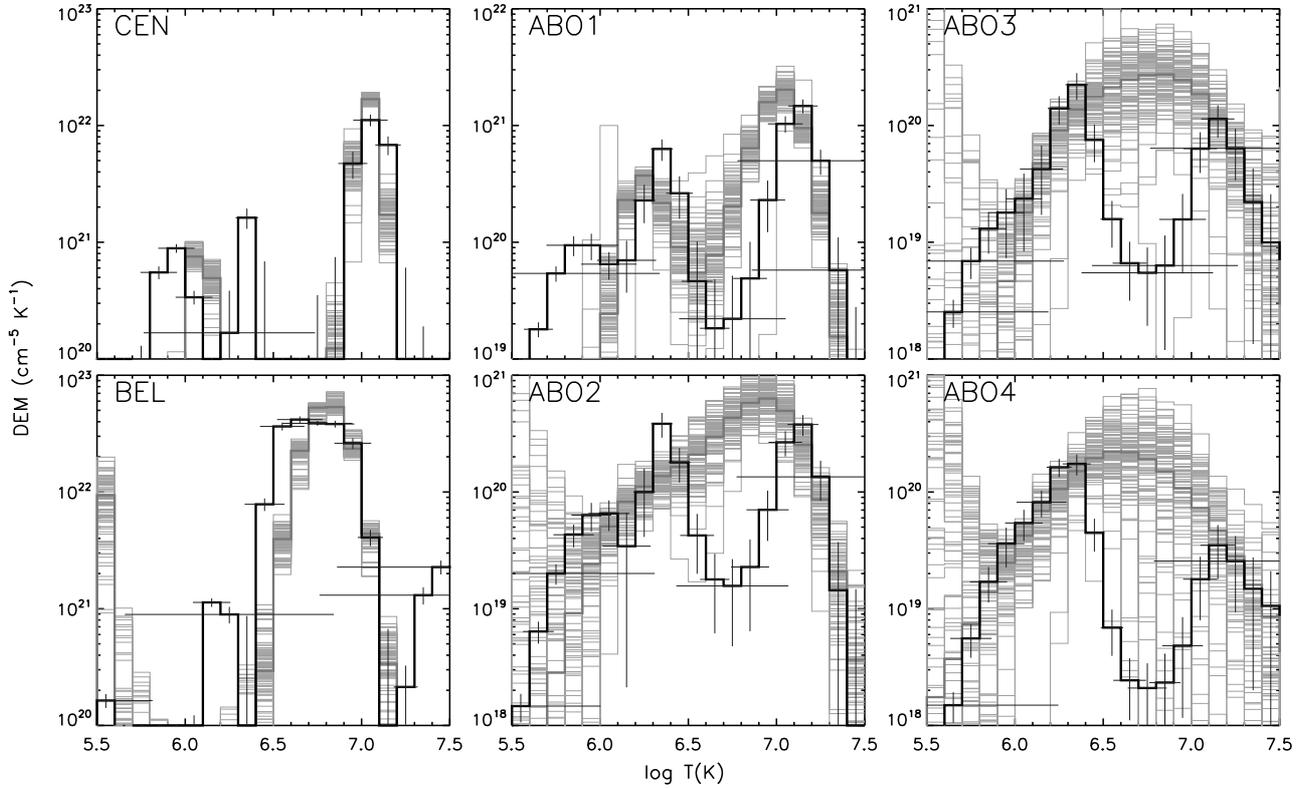}
\caption{DEM distributions for six small regions (5 by 5 pixels) of the SOL2011-10-22T11:10 flare at 12:35~UT (see Fig.~\ref{fig8}). The DEMs obtained with the use of the method-1 are shown by the gray histograms (basic DEM -- thick line, 100 Monte Carlo realizations representing estimation of errors -- thin lines). The DEMs calculated using the method-2 are represented by the black histogram with error bars.}
\label{fig9}
\end{figure*}

To study temperature distribution in more details we calculated DEM for six selected small regions. The regions are located within CXS and above it (see Fig.~\ref{fig8}). For each region DEM distributions obtained from method-1 and method-2 are shown in Fig.~\ref{fig9}.

The CEN region is located at the CXS centroid (see Fig.~\ref{fig8}). Both methods show consistent results for this region. There are two components of DEM. The first one, around 1~MK, is probably connected to quiet corona in the line of sight. Hot plasma located at the CXS centroid is responsible for the second component of DEM, which peaks at 10.0-12.6~MK. Both methods show the hot component as a narrow one -- most of the plasma is in the temperature range 7.9-15.8~MK. 

The second region (labeled BEL) is located below the CXS centroid. There is one reliable component in the DEM distribution, i.e. the results from both methods are similar and errors in each method are small. The BEL region is cooler than the CEN -- its DEM peak centers at 5.6-6.3~MK. However, the DEM is broad and includes plasma at the temperature 4-10~MK. The high temperature component of DEM (above 13~MK) obtained from the method-2 is not reliable because of errors (see large error bars). Comparison of DEM distributions for the CEN and BEL regions gives a similar conclusion as the obtained from the inspection of DEM maps. Regions located within CXS at a lower altitude than the center of the CXS are colder than in the center.

Other four regions (labeled ABO1/2/3/4, see Fig.~\ref{fig8}) are located above the CXS centroid. The ABO1 area is near the upper edge of the CXS defined by intensity isoline 0.5 (with respect to the brightest pixel of the CXS). There are some differences between the DEM distributions for this region obtained from both applied method. The method-1 gives two components in DEM while method-2 gives three. The weakest and coolest maximum (at T<1~MK) is present only in the method-2 DEM. The other two components are at similar positions in both methods. However, the components in the method-1 DEM are shifted towards a lower temperature by 0.1 dex. The hottest component is dominant and peaks at 12.6-15.8~MK (method-2) or 10.0-12.6~MK (method-1). DEMs distributions for the ABO1 fall off at high temperatures more slowly than for the CEN. This indicates that plasma in the ABO1 is on average somewhat hotter than in the CEN.

In the case of the ABO2, ABO3 and ABO4 regions DEM distributions obtained from the method-1 and the method-2 differs significantly. The method-1 gives the very broad DEMs with one maximum. The method-2 DEMs are similarly broad but have two distinct components. \citet{hannah2012} showed that for AIA data the method-2 performs much better than method-1 in the case of DEM distributions with two or more components. If separation of the components is smaller than 1 dex, the method-1 could not be able to reproduce the components. In the case of similarly strong components the method-1 may produce broad DEM encompassing both components but with one maximum located somewhere between original maxima. If one component is significantly weaker than the other one then in DEM recovered by the method-1 the weaker component may be missing. This can explain differences between DEMs obtained for the ABO1 region -- two versus three components.

Given the conclusions presented by \citet{hannah2012} we rely on DEMs recovered for the ABO2, ABO3 and ABO4 regions using the method-2. For each of these three regions DEM consists of two components. The cooler component peaks at about 2~MK. The second component, more important in our analysis, reaches the maximal temperature at 12.6-15.8~MK in all three regions. This is similar to the DEM of the ABO1 region. The key difference between all ABO regions is that for the ABO3 and ABO4, located at the highest altitudes, DEMs fall off noticeably more slowly at high temperatures. Thus, a relative abundance of plasma above 15.8~MK is probably higher in these two regions. However, it should be pointed out that errors in DEMs at the highest temperatures are not small.

It is worth to note that two components of DEM for the SAHR seem to be a general property of supra-arcade regions. In \citet{hanneman2014} 4 flares with a supra-arcade region were analyzed including the SOL2011-10-22T11:10 flare. The authors focused their attention on plasma characteristics of supra-arcade downflows with the respect to a surrounding supra-arcade region. In most of analyzed by them cases supra-arcade regions show two components in DEM distribution -- warm ($\approx2$~MK) and hot ($\approx10$~MK). Similar result for the SOL2011-10-22T11:10 flare was obtained by \citet{scott2016}. Authors interpret the result as follows. Warm component is due to plasma of surrounding quiet corona along the line of sight. Hot component is emitted by plasma in the supra-arcade region.

Comparison of DEMs for the six selected regions shows that there is a temperature gradient in the supra-arcade hot region. A peak temperature in the DEMs shifts from 5.6-6.3~MK at the lower edge of the CXS to 12.6-15.8~MK at its upper edge. A peak of DEMs above the CXS stays at 12.6-15.8~MK. However, a relative abundance of plasma hotter than 15.8~MK probably increases slightly with altitude.

An analysis of DEMs shows that the CXS was located in the lower part of the SAHR where emission measure was the highest. However, this part of the region is not the hottest. Taking into account the uncertainties in the high-temperature end of DEMs we can conclude that the hottest part of the SAHR ranges from the upper edge of the CXS up. The \textit{RHESSI} source did not extend higher because of significantly lower emission measure of plasma. A dynamic range of images reconstructed from \textit{RHESSI} data is usually about 10:1. A dynamic range in emission measure of the SAHR is much greater, i.e. about 1000:1 (see DEMs in Fig.~\ref{fig9}).

\section{Discussion and conclusions}
\label{sec-dis}

Having analyzed data from AIA and \textit{RHESSI}, let us track the SOL2011-10-22T11:10 flare evolution with an emphasis on the coronal X-ray source (CXS). At the flare beginning the eruption was recorded by AIA at about 10:00~UT. Assuming the standard flare model, a current sheet with the magnetic reconnection formed below the eruption. Loops formed in the reconnection region shrank rapidly and they were visible as bright supra-arcade downflowing loops (SADL, \citet{savage2011, savage2012}) till about 12:00~UT.

Before and during the maximum of the flare it consisted of a set of bright and hot loops (HLs, $T\approx10$~MK). Later the loops cooled down and reappeared as warm post-reconnection flare loops (PFLs, $T\approx 1$~MK). The PFLs were accompanied by a bright supra-arcade hot region (SAHR, $T\approx10$~MK) that was located above. In the SAHR there were visible supra-arcade downflows (SADs).

The evolution of the coronal X-ray source recorded by \textit{RHESSI} in X-rays can be divided into two main stages. The first stage includes the rise and maximum of the flare, i.e. the first four time intervals of those selected from \textit{RHESSI} data. During that stage the CXS was located at or very near the bright apexes of the HLs. The transition from the HLs to the PFLs with the SAHR was not well observed by \textit{RHESSI} due to the satellite night. The second stage of the CXS evolution started about 12:00~UT and lasted at least till 16:00~UT, i.e. it included the flare decay phase. During that stage the CXS was embedded in the lower part of the SAHR, just above the PFLs.

Although the stage 1 and the transition to stage 2 are interesting from the point of view of understanding coronal sources, in this paper we gave our attention to the stage 2. Coronal sources as observed in X-rays are characterized by slow and gradual evolution, smooth and diffuse appearance and stamina lasting many hours. These features are exceptionally conspicuous and surprising during the decay phase of long-duration events (LDEs) that may last for hours on end. In this paper we wanted to find out what is the relation between coronal sources and structures observed usually in EUV during the decay phase of LDEs. This may shed light on the ``mystery'' of the sources.

The CXS in the stage 2 existed as long as the SAHR. Moreover, when the SAHR divided into two parts also the CXS did it. Thus, a strong relation between the CXS and the SAHR is evident. If the CXS owes its existence and physical characteristics to the SAHR, the region needs our attention. A bright curtain of plasma rising above an arcade of post-reconnection flare loops, often observed in large flares, is supposed to be a region of almost vertical magnetic field \citep[]{mckenzie1999, hanneman2014}. According to the standard flare model such a vertical region is formed below an eruption/coronal mass ejection and separates anti-parallel magnetic field lines inflowing into the reconnection zone \citep[]{yokoyama1997, yokoyama2001}. A few mechanisms had been proposed to explain enhanced plasma emission form this supra-arcade region \citep[]{seaton2009, reeves2010, scott2013}. Plasma visible as SAHR may be contained within a current sheet itself and/or within a \textit{thermal halo}, although the current sheet itself may be too thin to enable a significant emission measure. The halo abuts the current sheet from both sides. \citet{scott2013} presented a model that can explain how a thermal halo is formed. Flux tubes formed in reconnection retract rapidly down towards PFLs. The tubes act as a peristaltic pump on unreconnected magnetic field lines that surround a current sheet. As a result slow magnetosonic shocks can develop in the disturbed field lines. The authors argue that shocks may reach the chromosphere causing evaporation. This evaporation fills the region of thermal halo with dense and hot plasma. Plasma in the thermal halo can be also heated by radiation emitted by plasma within a current sheet. \citet{seaton2009} presented a different model where a thermal halo is formed through thermal conduction from a current sheet into surrounding plasma and then down along unreconnected field to the chromosphere entailing evaporation.

According to the interpretation of supra-arcade downflows (SADs) given by \citet{savage2012} based on an analysis of the SOL2011-10-22T11:10 flare, retracting loops create behind the regions of depleted density (SADs) while moving in plasma of the SAHR. The speed of the SADs decreased from about $100$~km s$^{-1}$ to about $2$~km s$^{-1}$ while approaching the tops of the PFLs. This deceleration could be attributed (in part) to drag force exerted on the retracting loops by pressure of plasma in the SAHR (Scott et al. 2013). If this drag force is actually present, it could be additional source of heating of plasma in the SAHR.

Apart of details of the SAHR formation, important thing is a relative position and size of the CXS and the SAHR. One of the observational facts is that in the analyzed flare the CXS was considerably smaller than the SAHR. The CXS was located in the lower part of the SAHR. Our analysis of a differential emission measure (DEM) for the selected time 12:35~UT shows that there was a temperature gradient in the SAHR -- the higher the altitude the higher the temperature. The coolest plasma ($6.3-7.9$~MK) was located just above the PFLs, the hottest ($12.6-15.8$~MK) -- at the altitude of the upper edge of the CXS and higher. Above the CXS the SAHR was almost uniformly hot (DEM peak at $12.6-15.8$~MK) with possibly larger relative abundance of plasma hotter than 15.8~MK. The CXS was co-aligned with a part of the SAHR where an emission measure in temperature range $6.3-15.8$~MK was the highest. This temperature range corresponds well to the temperatures obtained for the CXS from imaging spectrometry -- two thermal components at 8.6 and 13.9~MK (see Table~\ref{tbl-1}). We compared also emission measures obtained from both instruments for the same area defined by the intensity isoline 0.5 with respect to the brightest pixel of the CXS. The emission measure of the CXS ($EM_{\rm CXS} \approx 8.8 \times 10^{48} {\rm~cm}^{-3}$) is in a good agreement with the emission measure of the co-aligned part of the SAHR for temperature range $6.3-15.8$~MK ($EM_{\rm SAHR} \approx 9.3 \times 10^{48} {\rm~cm}^{-3}$). A significantly lower emission measure of the plasma in the upper part of the SAHR and a limited dynamic range of the \textit{RHESSI} images are the most likely causes for which the CXS did not extended to the entire SAHR.

We can conclude that X-ray emission recorded by \textit{RHESSI}, visible as the CXS came from the part of the SAHR that had the highest emission measure and simultaneously the temperature within the range of \textit{RHESSI} thermal-response ($\gtrsim7$~MK). However, there is one important difference between the CXS and the co-aligned part of the SAHR. The SAHR was a dynamic region consisting of small-scale structures (Fig.~\ref{fig4}, see also \citet{mckenzie2013} and \citet{innes2014} for more information on dynamical structure of supra-arcade regions). Whereas the CXS seemed to be smooth, structureless. We run several simulations using real and synthetic \textit{RHESSI} data, but we did not find any strong evidence that the CXS had a small-scale structure corresponding to the structure of the SAHR. However, it can not be excluded that this situation results from instrumental (\textit{RHESSI}) limitations. We obtained an unexpected result of our simulation that may have serious consequences.  In some configuration of X-ray sources, i.e. a large source closely accompanied by a much smaller source(s), the large source may be lost in a image reconstruction process. Such an artificial lack of some sources in the reconstructed images is disturbing as it may lead to wrong conclusions. The issue should be addressed from the point of view of image reconstruction algorithms.

Physical parameters (temperature, T, emission measure, EM) of the CXS obtained from \textit{RHESSI} imaging spectroscopy showed slow time-variations. During the decay phase the source cooled and fainted very slowly. For at least almost five hours after the maximum of the flare a temperature of the CXS was close to 10~MK. The SAHR was observable in the 94~{\AA} and 131~{\AA} AIA bands for similarly long period of time. As hot and dense plasma cools very fast, the CXS (SAHR) needed a constant energy supply. The CXS without such an energy supply should cool down in a few hundreds of seconds. To study energy demand of the CXS we calculated its energy balance. Three cooling processes were included in this balance: conduction, radiation, and plasma inflow/outflow.

\begin{equation}
\label{e1}
\left (\frac{d{E_{th}}}{dt}\right )_{\rm obs} = \left( \frac{d{E_{th}}}{dt} \right)_{\rm flow} - L_{\rm c} - L_{\rm r} + H ,
\end{equation}

where:

\begin{itemize}
	
	\item ${E_{\rm th}} = 3nk_{\rm B} TV$ is total thermal energy of the CXS ($n, T, V$ - are an electron density, a temperature and a volume of the CXS, respectively). The electron density is calculated as $n=(EM/V)^{1/2}$ ($EM$ is emission measure). To calculate the volume we used the cross-section area, $A$, of the CXS from \textit{RHESSI} images and the line-of-sight size ($d = 5 \times 10^{8} {\rm~cm}$) taken from \citet{savage2010}. In the paper the authors estimated a thickness of a current sheet (a SAHR) in a long-duration flare observed edge-on. We assumed that a line-of-sight size of the CXS is the same as a thickness of the SAHR and similar to the thickness measured by the mentioned authors. The temperature, the emission measure, and the cross-section area of the CXS are given in Table~\ref{tbl-1}.
	
	\item $\left (\frac{d{E_{th}}}{dt}\right )_{\rm obs}$ is the rate of change of ${E_{\rm th}}$ estimated from the observed parameters of the CXS.

	\item $\left (\frac{d{E_{th}}}{dt}\right )_{\rm flow}$ is the rate of change of ${E_{\rm th}}$ due to the plasma inflow/outflow into/from the source. This term was calculated as $\left (\frac{d{E_{th}}}{dt}\right )_{\rm flow} = 3k_{\rm B} TV\left (\frac{dn}{dt}\right)$.

	\item $L_{\rm c}$ is the non-local conductive cooling rate, $L_C = q_{\rm nl} \times L_{\rm c,S}$ where $L_{\rm c,S}$ is the Spitzer conductive cooling rate \citep[]{spitzer1962}. It has been shown that when the mean free path of thermal electrons is larger than $0.1-1$\% of a temperature variation length scale, conductivity becomes non-local and lower than the Spitzer conductivity \citep[]{luciani1983, campbell1984}. Such conditions are not rare in solar flares. The value of the coefficient $q_{nl}$ describing how much the non-local conductivity is smaller than the Spitzer one we took from \citet{campbell1984}. The Spitzer conductive cooling rate was calculated as $L_{\rm c,S} = 9.2 \times 10^{-7} T^{5/2} \nabla T A$ with the approximation $\nabla T \propto T/h$, where $h$ is an altitude of the CXS's centroid above the photosphere measured in \textit{RHESSI} images (see Table~\ref{tbl-1}).

	\item $L_{\rm r}$ is the radiative cooling rate, $L_{\rm r} = EM \times \Phi(T)$ where $\Phi(T)$ is the radiative loss function taken from \citet[]{colgan2008} and \citet[]{dere2009}.
	
	\item $H$ is the heating rate or thermal energy input rate.
	
\end{itemize}

\begin{figure}
\resizebox{\hsize}{!}{\includegraphics{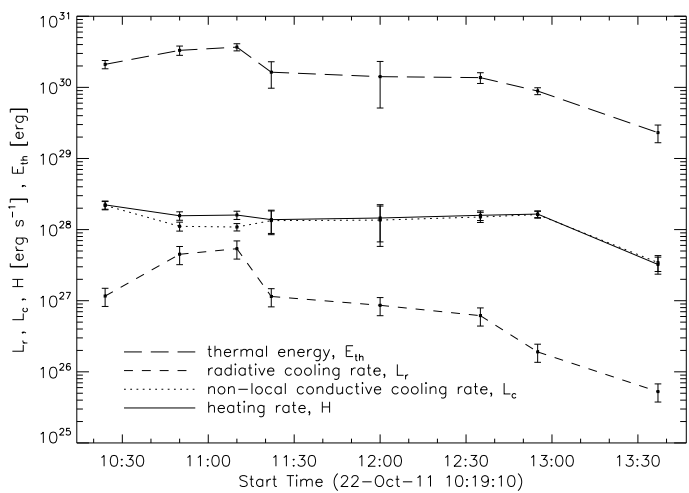}}
\caption{Components of the energy balance (see Eq.~\ref{e1}) calculated for the coronal X-ray source of the SOL2011-10-22T11:10 flare.}
\label{fig10}
\end{figure}

Using Eq.\ref{e1} and parameters of the CXS given in Table~\ref{tbl-1} the heating rate, $H$, can be estimated. The results are shown in Fig.~\ref{fig10}. For more than 3 hours (till 13:37~UT) $H$ was in the range $3-20 \times 10^{27} {\rm~erg} {\rm~s}^{-1}$, i.e. the CXS was constantly heated. The total amount of thermal energy added to the CXS is estimated as $E^{\rm tot}_{\rm th} = 1.5 - 1.9 \times 10^{32} {\rm~erg}$ of which about 65\% during the decay phase. This result should be treated as the lower limit because the CXS was observable till 16:00~UT (see Fig.~\ref{fig2}). Unfortunately, we were not able to obtain $T$ and $EM$ from imaging spectroscopy for time later than 13:37~UT. Assuming that from 13:37~UT a heating rate, $H$, linearly decreased to zero at 16:00~UT, the total thermal energy would be $E^{\rm tot}_{\rm th} = 1.6 - 2.0 \times 10^{32} {\rm~erg}$. The SAHR is bigger than the CXS thus, the thermal energy needed to sustain the whole supra-arcade region of hot plasma had to be larger than the estimated for the CXS. The energy demand of the CXS indicates that energy release was ongoing during the whole evolution of the analyzed flare, even long after its maximum.

Another piece of evidence of ongoing energy release are supra-arcade downflows. SADs are considered to be a manifestation of magnetic reconnection \citep[]{asai2004, khan2007}. SADs were streaming down uninterruptedly during the decay phase of the SOL2011-10-22T11:10 flare till about 15:00~UT (see Fig.~\ref{fig7}). A constant stream of SADs suggests constant reconnection and thus a steady source of thermal energy for the supra-arcade region (CXS, SAHR) enabling a slow gradual evolution of the region. Just after 15:00~UT the SADs became virtually invisible. The SAHR and the CXS vanished not much later, about $16-17$~UT. Apparently the magnetic reconnection became too weak at that time to support an existence of hot and dense plasma, but it did not terminate completely. New post-reconnection flare loops appeared even at 20:00~UT.

\citet{vorpahl1977} based on X-ray observations taken from \textit{Skylab} concluded that to explain the persistence of CXSs not only heating but also some restriction mechanism, efficiently preventing an outflow of mass and energy from the sources, is needed. Similar suggestion was presented later, after the analysis of the \textit{RHESSI} and the \textit{Yohkoh} data \citep[e.g.]{jiang2006, kolomanski2007}. We tried to verify this suggestion using a hydrodynamical model of magnetic loops.

In our model we assumed that SADs are caused by loops retracting downward through a supra-arcade hot region as proposed by e.g. \citet{savage2012}. The retracting loops (RLs) decelerated and accumulated in the lower part of the SAHR where the CXS was observed. Thus, the RLs were embedded in the CXS, piercing through it. \citet{hanneman2014} showed that physical parameters of SADs are coupled with parameters of an adjacent part of SAHR -- the temperature of SADs is similar to the temperature of SAHR, while their emission measure (density) is slightly lower. Thus, if any restriction of an energy outflow worked in the CXS, it should probably influence cooling of the retracting loops.

We utilized the Palermo-Harvard (PH) hydrodynamical model to study the cooling of RLs. The PH model is 1D model which enables to analyze a hydrodynamical response of a loop to heating. For a detailed description of the PH model see \citet{peres1982} and \citet{betta1997}. We performed simulations of a thermal evolution of a flaring loop with a semi-length of $L=9\times10^{9}$~cm. This was approximately semi-length of the RLs having apexes near the centroid of the CXS at 12:35~UT.

The level of the thermal energy release was selected high enough to allow plasma to reach the temperature of $10-15$~MK and the density of a $4-9\times10^{9}$ cm$^{-3}$ in the top of the modeled loop. These values of $T$ and $N$ were selected to be in the agreement with the results of DEM analysis for the CEN region (see the Section 3.3). The region was a part of the SAHR located at the centroid of the CXS.

Firstly, the modeled loop was heated for a few hundreds of seconds enabling to reach the required $T$ and $N$. After that the heating was abruptly switched off and the loop was allowed to cool down. The was no restriction imposed on mass and energy outflow.  It took $\approx3500-5500$~s for the modeled loops to cool down from $10-15$~MK to $1$~MK. The modeled cooling time is similar to the observed one. The tops of RLs that were near the centroid of the CXS at 12:35~UT, appeared as the PFLs ($T\approx1$~MK) after about $5000$~s (see Fig.~\ref{fig7}). Thus, the numerical modeling suggests that during the cooling of the RLs there was no significant restriction on energy and mass outflow. We conclude that probably such a restriction was not present in the case of the coronal source of the SOL2011-10-22T11:10 flare. However, it can not be excluded that a restriction mechanism works only in a supra-arcade hot region and does not influence retracting loops (SADs). A further study of SADs and SAHR/CXS is needed to fully understand how they interact.

Simultaneous observations of solar flares performed by AIA and \textit{RHESSI} can give a unique opportunity to study coronal sources -- their structure, parameters, evolution and relation between structures observed in EUV and X-rays. But still there are limitations that impede a detailed analysis and leave some questions open. One of such questions is a small-scale structure of coronal sources as observed in X-rays. \textit{RHESSI} is a great instrument but in this case it seams not to be sufficient. Observations with better spatial and temporal resolutions are needed. In the nearest future such observations will be provided by STIX (Spectrometer/Telescope for Imaging X-rays, \citet{benz2012}) onboard the \textit{Solar Orbiter} spacecraft. A launch of this ESA mission is targeted for October 2018. \textit{Solar Orbiter} will be put on a highly eccentric orbit around the Sun with the perihelion at only 0.28 AU. Observing the Sun with STIX from such a small distance is a chance to obtain images and spectroscopy of X-ray emission of solar flares good enough for better understanding of coronal sources. Studies of the sources are one of the key factors in our understanding of solar and stellar flares.

\begin{acknowledgements}
Authors acknowledge financial support from the Polish National Science Centre grants 2011/03/B/ST9/00104 and 2011/01/M/ST9/06096.
We thanks Prof. Fabio Reale from Universita di Palermo for enabling us to use Palermo-Harvard hydrodynamic code. 
Calculations have been carried out using resources provided by Wroclaw Centre for Networking and Supercomputing (http://wcss.pl), grant No. 221.
\end{acknowledgements}


\begin{thebibliography}{}


\bibitem[Asai et al.(2004)]{asai2004} 
Asai, A., Yokoyama, T., Shimojo, M., \& Shibata, K.\ 2004, ApJL, 605, L77

\bibitem[Acton et al.(1992)]{acton1992} 
Acton, L.~W., Feldman, U., Bruner, M.~E., et al.\ 1992, \pasj, 44, L71 

\bibitem[Aschwanden et al.(2004)]{aschwanden2004} 
Aschwanden, M.~J., Metcalf, T.~R., Krucker, S., et al.\ 2004, \solphys, 219, 149 

\bibitem[B\c{a}k-St\c{e}\'slicka et al.(2011)]{baksteslicka2011} 
B\c{a}k-St\c{e}\'slicka, U., Mrozek, T., \& Ko{\l}oma{\'n}ski, S.\ 2011, \solphys, 271, 75 

\bibitem[Benz et al(2012)]{benz2012}
Benz, A.~O., Krucker, S., Hurford, G.~J., et al.\ 2012, \procspie, 8443, 84433L 

\bibitem[Bertero et al.(1985)]{bertero1985} 
Bertero, M., DeMol, C., \& Pike, E.~R.\ 1985, Inverse Problems, 1, 301

\bibitem[Betta et al.(1997)]{betta1997} 
Betta, R., Peres, G., Reale, F., \& Serio, S.\ 1997, \aaps, 122, 585 

\bibitem[Boerner et al.(2012)]{boerner2012} 
Boerner, P., Edwards, C., Lemen, J., et al.\ 2012, \solphys, 275, 41 

\bibitem[Campbell(1984)]{campbell1984}
Campbell, P.M.,\ 1984, Physical Review A, 30, 365

\bibitem[Caspi \& Lin(2010)]{caspi2010} 
Caspi, A., \& Lin, R.~P.\ 2010, \apjl, 725, L161 

\bibitem[Cheng et al.(2012)]{cheng2012} 
Cheng, X., Zhang, J., Saar, S.~H., \& Ding, M.~D.\ 2012, \apj, 761, 62

\bibitem[Colgan et al.(2008)]{colgan2008}
Colgan, J., Abdallah, Jr.~J., Sherrill, M.~E.\ 2008, \apj, 689, 585

\bibitem[Craig \& Brown(1986)]{craig1986} 
Craig, I.~J.~D. \& Brown, J.~C.\ 1986, Inverse problems in astronomy (Bristol: Hilger)

\bibitem[Culhane et al.(2007)]{culhane2007} 
Culhane, J.~L., Harra, L.~K., James, A.~M., et al.\ 2007, \solphys, 243, 19 

\bibitem[Dennis \& Pernak(2009)]{dennis2009} 
Dennis, B~R. \& Pernak, R.~L.\ 2009, \apj, 698, 2131 

\bibitem[Dere et al.(2009)]{dere2009}
Dere, K.~P., Landi, E., Young P.~R., et al.\ 2009, \aap, 498, 915

\bibitem[Donnelly et al.(1977)]{donnelly1977}
Donnelly, R.~F., Grubb, R.~N., \& Cowley, F.~C.\ 1977, NOAA Tech. Memo. ERL SEL-48

\bibitem[Doschek \& Feldman(1996)]{doschek1996} 
Doschek, G.~A., \& Feldman, U.\ 1996, \apj, 459, 773 

\bibitem[Doschek et al.(1995)]{doschek1995} 
Doschek, G.~A., Strong, K.~T., \& Tsuneta, S.\ 1995, \apj, 440, 370 

\bibitem[Feldman et al.(1995)]{feldman1995} 
Feldman, U., Seely, J.~F., Doschek, G.~A., et al.\ 1995, \apj, 446, 860 

\bibitem[Freeland \& Handy(1998)]{freeland1998}
Freeland, S.~L., \& Handy, B.~N.\ 1998, \solphys, 182, 497

\bibitem[Golub at al.(2004)]{golub2004} 
Golub, L., Deluca, E.~E., Sette, A., \& Weber, M.\ 2004, 
in The Solar-B Mission and the Forefront of Solar Physics, ed. T. Sakurai, \& T. Sekii, ASP Conf. Ser., 325, 217

\bibitem[Handy et al.(1999)]{handy1999} 
Handy, B.~N., Acton, L.~W., Kankelborg, C.~C., et al.\ 1999, \solphys, 187, 229 

\bibitem[Hannah \& Kontar(2012)]{hannah2012} 
Hannah, I.~G., \& Kontar, E.~P.\ 2012, \aap, 539, A146

\bibitem[Hanneman \& Reeves(2014)]{hanneman2014} 
Hanneman, W.~J., \& Reeves, K.~K.\ 2014, \apj, 786, 95

\bibitem[Hudson \& McKenzie(2000)]{hudson2000} 
Hudson, H.~S., \& McKenzie, D.~E.\ 2000, High Energy Solar Physics Workshop - Anticipating Hess!, 206, 221 

\bibitem[Hurford et al.(2002)]{hurford2002}
Hurford, G.~J., Schmahl, E.~J., Schwartz, R.~A., et al.\ 2002, \solphys, 210, 61 

\bibitem[Innes et al.(2014)]{innes2014}
Innes, D.~E., Guo, L.-J., Bhattacharjee, A., Huang, Y.-M., \& Schmit, D.\ 2014, \apj, 796, 27

\bibitem[Jakimiec et al.(1998)]{jakimiec1998} 
Jakimiec, J., Tomczak, M., Falewicz, R., Phillips, K.~J.~H., \& Fludra, A.\ 1998, \aap, 334, 1112 

\bibitem[Jiang et al.(2006)]{jiang2006} 
Jiang, Y.~W., Liu, S., Liu, W., \& Petrosian, V.\ 2006, \apj, 638, 1140

\bibitem[Kahler(1977)]{kahler1977} 
Kahler, S.\ 1977, \apj, 214, 891 

\bibitem[Khan et al.(2007)]{khan2007} 
Khan, J.~I., Bain, H.~M., \& Fletcher, L.\ 2007, \aap, 475, 333

\bibitem[Ko{\l}oma{\'n}ski(2007)]{kolomanski2007} 
Ko{\l}oma{\'n}ski, S.\ 2007, \aap, 465, 1035 

\bibitem[Ko{\l}oma{\'n}ski et al.(2011)]{kolomanski2011} 
Ko{\l}oma{\'n}ski, S., Mrozek, T., \& B{\c a}k-St{\c e}{\'s}licka, U.\ 2011, \aap, 531, A57 

\bibitem[Kosugi et al.(2007)]{kosugi2007}
Kosugi, T., Matsuzaki, K., Sakao, T., et al.\ 2007, \solphys, 243, 3 

\bibitem[Krucker et al.(2008)]{krucker2008}
Krucker, S., Battaglia, M., Cargill, P.~J., et al.\ 2008, \aapr, 16, 155 

\bibitem[Lemen et al.(2012)]{lemen2012} 
Lemen, J.~R., Title, A.~M., Akin, D.~J., et al.\ 2012, \solphys, 275, 17 

\bibitem[Lin et al.(2003)]{lin2003} 
Lin, R.~P., Dennis, B.~R., \& Benz, A.~O.\ 2003, The Reuven Ramaty High-Energy Solar Spectroscopic Imager (RHESSI) - Mission Description and Early Results.~Edited by Robert P.~Lin, Brian R.~Dennis, \& Arnold O.~Benz.~ Reprinted from  Solar Physics, Volume 210, Nos.~1-2 (2002)  Kluwer Academic Publishers, Dordrecht

\bibitem[Lin et al.(2002)]{lin2002} 
Lin, R.~P., Dennis, B.~R., Hurford, G.~J., et al.\ 2002, \solphys, 210, 3 

\bibitem[Liu et al.(2013)]{liu2013}
Liu, W., Chen, Q., \& Petrosian, V.\ 2013, \apj, 767, 168

\bibitem[Longcope et al.(2010)]{longcope2010} 
Longcope, D.~W., Des Jardins, A.~C., Carranza-Fulmer, T., \& Qiu, J.\ 2010, \solphys, 267, 107 

\bibitem[Luciani et al.(1983)]{luciani1983} 
Luciani, J.~F., Mora, P., \& Virmont, J.\ 1983, Phys. Rev. Lett., 51, 1664

\bibitem[McKenzie(2000)]{mckenzie2000} 
McKenzie, D.~E.\ 2000, \solphys, 195, 381 

\bibitem[McKenzie(2013)]{mckenzie2013}
McKenzie, D.~E.\ 2013, \apj, 766, 39 

\bibitem[McKenzie \& Hudson(1999)]{mckenzie1999} 
McKenzie, D.~E., \& Hudson, H.~S.\ 1999, \apjl, 519, L93 

\bibitem[O'Dwyer et al.(2010)]{odwyer2010}
O'Dwyer, B., Del Zanna, G., Mason, H.~E., Weber, M.~A. \& Tripathi, D.\ 2010, \aap, 521, A21 

\bibitem[Ogawara et al.(1991)]{ogawara1991}
Ogawara, Y., Takano, T., Kato, T., et al.\ 1991, \solphys, 136, 1

\bibitem[Peres et al.(1982)]{peres1982} 
Peres, G., Serio, S., Vaiana, G.~S., \& Rosner, R.\ 1982, \apj, 252, 791 

\bibitem[Pesnell et al.(2012)]{pesnell2012} 
Pesnell, W.~D., Thompson, B.~J., \& Chamberlin, P.~C.\ 2012, \solphys, 275, 3 

\bibitem[Phillips et al.(2005)]{phillips2005} 
Phillips, K.~J.~H., Chifor, C., \& Landi, E.\ 2005, \apj, 626, 1110 

\bibitem[Pi\~na \& Puetter(1993)]{pina1993} 
Pi\~na, R.~K., \& Puetter, R.~C.\ 1993, \pasp, 105, 630 

\bibitem[Prato et al.(2006)]{prato2006} 
Prato, M., Piana, M., Brown, J.~C., et al.\ 2006, \solphys, 237, 61

\bibitem[Pre{\'s} \& Ko{\l}oma{\'n}ski(2009)]{pres2009} 
Pre{\'s}, P., \& Ko{\l}oma{\'n}ski, S.\ 2009, Central European Astrophysical Bulletin, 33, 233 

\bibitem[Priest \& Forbes(2000)]{priest2000} 
Priest, E., \& Forbes, T.\ 2000, Magnetic Reconnection, by Eric Priest and Terry Forbes, 
pp.~612.~ISBN 0521481791.~Cambridge, UK: Cambridge University Press, June 2000., 612 

\bibitem[Reeves et al.(2010)]{reeves2010} 
Reeves, K.~K., Linker, J.~A., Miki{\'c}, Z., \& Forbes, T.~G.\ 2010, \apj, 721, 1547

\bibitem[Savage et al.(2010)]{savage2010} 
Savage, S.~L., McKenzie, D.~E., Reeves, K.~K., Forbes, T.~G., \& Longcope, D.~W.\ 2010, \apj, 722, 329

\bibitem[Savage \& Mckenzie(2011)]{savage2011} 
Savage, S.~L., \& McKenzie, D.~E.\ 2011, \apj, 730, 98

\bibitem[Savage et al.(2012)]{savage2012} 
Savage, S.~L., McKenzie, D.~E., \& Reeves, K.~K.\ 2012, \apjl, 747, L40 

\bibitem[Seaton \& Forbes(2009)]{seaton2009} 
Seaton, D.~B., \& Forbes, T.~G.\ 2009, \apj, 701, 348

\bibitem[Scott et al.(2013)]{scott2013} 
Scott, R.~B., Longcope, D.~W., \& McKenzie, D.~E.,\ 2013, \apj, 776, 54

\bibitem[Scott et al.(2016)]{scott2016}
Scott, R.~B., McKenzie, D.~E., \& Longcope, D.~W.\ 2016, \apj, 819, 56 

\bibitem[Schmahl et al.(2007)]{schmahl2007}	
Schmahl, E.~J., Pernak, R.~L., Hurford, G.~J., Lee, J., \& Bong, S.\ 2007, \solphys, 240, 241 

\bibitem[Schmitt et al.(1996)]{schmitt1996} 
Schmitt, J.~H.~M.~M., Drake, J.~J., Stern, R.~A., \& Haisch, B.~M.\ 1996, \apj, 457, 882

\bibitem[Sheeley et al.(2004)]{sheeley2004} 
Sheeley, N.~R., Jr., Warren, H.~P., \& Wang, Y.-M.\ 2004, \apj, 616, 1224 

\bibitem[Spitzer(1962)]{spitzer1962}
Spitzer, L.\ 1962, Physics of Fully Ionized Gases (New York: Wiley), chap. 5

\bibitem[{\v S}vestka(2007)]{svestka2007}
{\v S}vestka, Z.\ 2007, \solphys, 246, 393 

\bibitem[Tikhonov(1963)]{tikhonov1963} 
Tikhonov, A.~N.\ 1963, Soviet Math. Dokl., 4, 1035

\bibitem[Tsuneta et al.(1991)]{tsuneta1991}
Tsuneta, S., Acton, L., Bruner, M., et al.\ 1991, \solphys, 136, 37 

\bibitem[V{\"a}{\"a}n{\"a}nen \& Pohjolainen(2007)]{vaananen2007} 
V{\"a}{\"a}n{\"a}nen, M., \& Pohjolainen, S.\ 2007, \solphys, 241, 279 

\bibitem[Vorpahl et al.(1977)]{vorpahl1977} 
Vorpahl, J.~A., Tandberg-Hanssen, E., \& Smith, J.~B., Jr.\ 1977, \apj, 212, 550 

\bibitem[Warren et al.(1999)]{warren1999} 
Warren, H.~P., Bookbinder, J.~A., Forbes, T.~G., et al.\ 1999, \apjl, 527, L121 

\bibitem[Warren et al.(2011)]{warren2011}
Warren, H.~P., O'Brien, C.~M., \& Sheeley, N.~R., Jr.\ 2011, \apj, 742, 92

\bibitem[Weber et al.(2004)]{weber2004} 
Weber, M.~A., Deluca, E.~E.,Golub, L., \& Sette, A. L.\ 2004, 
in Multi-Wavelength Investigations of Solar Activity, ed. A.~V. Stepanov, E.~E. Benevolenskaya, \& A.~G. Kosovichev, IAU Symp., 223, 321

\bibitem[White et al.(2002)]{white2002} 
White, S.~M., Kundu, M.~R., Garaimov, V.~I., Yokoyama, T., \& Sato, J.\ 2002, \apj, 576, 505

\bibitem[Yokoyama \& Shibata(1997)]{yokoyama1997} 
Yokoyama, T., \& Shibata, K.\ 1997, \apj, 474, L61

\bibitem[Yokoyama \& Shibata(2001)]{yokoyama2001} 
Yokoyama, T., \& Shibata, K.\ 2001, \apj, 549, 1160


\end{thebibliography}
\end{document}